\documentclass[aps,prd,twocolumn,groupedaddress,showpacs,nofootinbib]{revtex4}

\usepackage{graphicx}
\usepackage{dcolumn}
\usepackage{bm}

\usepackage{color}
\usepackage{bm}

\definecolor{darkred}{rgb}{.8,0,0}

\definecolor{darkblue}{rgb}{0,0,.7}

\newcommand{\rerevision}{\cdgreen}
\definecolor{darkgreen}{rgb}{0,.8,0}
\newcommand{\cdgreen}{\color{darkgreen}}

\usepackage{amsmath, amsthm}
\def\ben{\begin{enumerate}} \def\een{\end{enumerate}}
\def\beq{\begin{equation}} \def\eeq{\end{equation}}
\def\beqn{\begin{equation*}} \def\eeqn{\end{equation*}}
\def\bea{\begin{eqnarray}} \def\eea{\end{eqnarray}}
\def\ba{\begin{array}} \def\ea{\end{array}}
\def\beann{\begin{eqnarray*}} \def\eeann{\end{eqnarray*}}
\def\beasn{\begin{sneqnarray}} \def\eeasn{\end{sneqnarray}}

\def\rr{\rerevision}

\begin{document}

\author{J. M. Pons}
\thanks{email: {\tt pons@ecm.ub.es}}
\affiliation{Departament d'Estructura i Constituents de la Mat\`eria and Institut de
Ci\`encies del Cosmos, Universitat de Barcelona,
Diagonal 647, 08028 Barcelona, Catalonia, Spain}
\author{D. C. Salisbury}
\thanks{E-mail:\tt{dsalisbury@austincollege.edu}}
\affiliation{Max-Planck-Institut f\"ur Wissenschaftsgeschichte,
Boltzmannstrasse 22,
14195 Berlin, Germany\\ Department of Physics,
Austin College, Sherman, Texas 75090-4440, USA}
\author{K. A. Sundermeyer}
\thanks{E-mail:\tt{ksun@gmx.de}}
\affiliation{Freie Universit\"at Berlin, Fachbereich Physik,
Institute for Theoretical Physics, Arnimallee 14, 14195 Berlin,
Germany}

\pacs{4.20.Fy, 4.60.Ds.}

\title{Revisiting observables in generally covariant
theories  in the light of gauge fixing methods
}

\pacs{4.20.Fy, 4.60.Ds.}

\begin{abstract}
We derive for generally covariant theories the generic dependency of
observables on the original fields, corresponding to
coordinate-dependent gauge fixings. This gauge choice is equivalent
to a choice of intrinsically defined coordinates accomplished with
the aid of spacetime scalar fields. With our approach we make full
contact with, and give a new perspective to, the ``evolving constants
of motion" program. We are able to directly derive generic
properties of observables, especially their dynamics and their
Poisson algebra in terms of Dirac brackets, extending earlier
results in the literature. We also give a new interpretation of the
observables as limits of canonical maps.

\end{abstract}

\maketitle
\section{Introduction}

Theories with gauge symmetries exhibit a mathematically redundant
description of the same physical setting. Gauge transformations,
defined in the space of field configurations, map solutions of the
equations of motion to other solutions that have the same physical
contents. Generally covariant theories - like general relativity -
and Yang-Mills theories are the most relevant examples of this type
of theories. The concept of observables for these theories has long
been a topic of discussion. We refer for instance to Bergmann's
particularly insightful treatment in general relativity
\cite{bergmann61}.

Although there exist different conceptions of observables in
theories with gauge symmetries, everyone in the community agrees
that these are quantities that are invariant under the respective
gauge transformations. Therefore they are sometimes also called
``gauge invariants", or simply ``invariants".  Together with that of
``observables", this is the terminology that will be employed here without making any distinction among them. 
Other language like ``complete observables", or ``Dirac observables" can be 
found in the literature\footnote{One may even distinguish Dirac from
Bergmann observables; see \cite{LP}.}. For generally covariant theories observables are
identified with those objects of the theory that are invariant under
coordinate transformations. They may be the classical versions of
quantum observables, although there is no a priori relation to
quantities that can be measured. In canonical versions of generally
covariant theories - understood  as the first step of canonical
quantizations -  it is more or less known from the work of
Rosenfeld, Dirac, and Bergmann that (1) the Hamiltonian is nothing
but a sum of constraints (relations between  fields and their canonically conjugate momenta)
defining a surface in the phase space,  (2) the local symmetries are
generated by gauge generators which can be expressed by the
constraints and (3) the observables are those objects which have
vanishing Poisson brackets on the constraint surface with the gauge
generators.

 In the preceding sentences we were simply referring to ``the
constraints". However, as known from the Rosenfeld-Dirac-Bergmann 
algorithm\footnote{Although every respectable review of ``Constrained
Dynamics" mentions the work of L.Rosenfeld it is not well-known in
the community which results he actually established in 1930, and
which later where re-established especially by Dirac. This will be
elaborated from a history-of-science point of view by one of the
authors (DCS) in a forthcoming publication in {\it Archive for History of Exact Science}. 
See also \cite{Salisbury:2007br,Salisbury:2009}.}, one must
distinguish first and second class constraints as well as various
generations of constraints (primary, secondary,...). Whereas the
notion of ``first/second" class is tied to the Poisson bracket
relations among the constraints, the ``generation classification"
depends on   the stage at which a given constraint appears when imposing 
consistency requirements on the equations of motion (EOM).

It is known from the (1+3) ADM-split that in generally covariant
theories the Hamiltonian  ${\cal H}_0$ constraint and the
momentum constraints ${\cal H}_a$ are secondary first-class
constraints. The Hamiltonian of the theory is built out of these constraints, with the possible addition of boundary terms, which may be necessary in order for the Hamiltonian to be a differentiable functional \cite{Regge:1974zd}; these terms have no effect on the dynamics, but they can be relevant as regards conserved quantities. There are also primary first-class constraints, namely
the vanishing momenta canonically conjugate to the lapse and shift
functions. We define as observables those quantities that
have weakly vanishing\footnote{ In Dirac's notation, ``weakly vanishing" means
vanishing on the constraint surface.} Poisson brackets with \underline{all} first
class constraints. (This definition is not entirely in agreement
with the definition of many others who only require weakly
vanishing Poisson brackets with the secondary first-class
constraints. Both definitions agree if one drops the lapse and shift
functions as canonical fields.)

Ultimately the interest in observables is due to the necessity of
identifying those quantities that can be predicted from a theory and
are subject to measurement. However, in the presence of phase-space
constraints this identification is not trivial.

In the case of Yang-Mills theories, with internal - i.e., non spacetime - gauge 
symmetries, observables are well known, and either local like the trace of the 
curvature $2$-form, or non-local like the Wilson loop. 
But, to this date, for generic general relativity no observables are
known. Only for some spacetimes with special asymptotic behavior or
additional Killing symmetries have observables been constructed;
this includes cosmologies and cylindrical waves. In these cases one
has been able to construct observables with the help of explicit
solutions of the respective field equations. Since the work of Torre
\cite{Torre:1993fq} one knows that the observables in general are
nonlocal, that is, functionals of the original fields and their
derivatives. Only recently did Dittrich \cite{dittrich07} and
Thiemann \cite{thiemann06} find a formal expression for these
functional invariants in the generic case, but for canonical
variables other than lapse and shift. We give an interpretation of
Dittrich's and Thiemann's expressions in terms of gauge choices and
intrinsic coordinates, where everything is based on the original
diffeomorphism symmetry of general relativity. This geometric and
physical route towards the observables furthermore permits us to
display some interesting properties of the functional invariants. In addition to the authors referred to above, we must mention the work by Lusanna and Pauri \cite{LP}.  They analyze the notion of observable, among other issues, from a somewhat different perspective.

 This article is written in the spirit of previous work
\cite{pss1997pr,ps04,shs07} essentially dealing with a deeper
understanding of the fate of the diffeomorphism group in the phase
space formulation of generally covariant theories.

In these articles \cite{pss1997pr,ps04,shs07} it was stressed that
\begin{itemize}
\item there is a maximal subgroup of the field-dependent
diffeomorphism group that can be realized as canonical transformations in
phase space. This subgroup of ``diffeomorphism-induced''
transformations can be characterized by asking for Legendre
projectability between the configuration-velocity and the phase space of the theory.

\item the lapse and the shift functions are not arbitrary
Lagrange multipliers in the Dirac Hamiltonian, but are canonical field variables.
Otherwise one is not able to realize the full group of 4-dimensional diffeomorphisms in phase
space. It is true, though, that the dynamics will relate the lapse and the shift
variables with the time derivatives of the Lagrange multipliers associated with the
primary constraints.

\item  a specific combination of all first class
constraints constitutes the generator of the diffeomorphism-induced
transformations on the original phase space. The Hamiltonian and
momentum constraints only generate transformations in a reduced
phase space - where the lapse and shift are turned into Lagrange
multipliers whereas their canonical momenta (the primary
constraints) are eliminated - and they are no longer related to the
original four-dimensional diffeomorphism group that included
transformations of the coordinate time.

\item time evolution in the phase space of generally covariant theories is distinguished
from gauge transformations.  On the other hand, 
projectability issues prevent the gauge group from containing time translations as a subgroup, 
in the sense that an element of the former
group effects the same time translation on all solution trajectories.

\item the observables may depend on  the coordinate time  of any observer, which may be quite arbitrary. Thus they are not necessarily constants of motion, although constants of motion can be extracted form them.

\item  from the perpective of gauge fixing methods, observables are nothing other than the full set of dynamical variables evaluated in an appropriately chosen intrinsic coordinate system. Equivalently, they may be obtained through a symmetry gauge transformation to the intrinsic coordinate system.

\end{itemize}

Notice that many of these findings are correlated by simply requiring
Legendre-projectabiliy. We also point
out that  some of them undermine folklore in the canonical
gravity community. In this paper we will elucidate further the final
bulleted point, and we will compare with other procedures for the
construction of observables that can be found in the literature.

Most of our results are proven to be valid locally. We do not address global issues in this article. Although we could asssume for the sake of simplicity that the spacetime is spatially compact without boundary this is not really relevant; we do not expect that the local construction of observables is changed by the topology of spacetime or  by possible extra terms in the Hamiltonian that appear, e.g., due to spatial non-compactness. Of course, as is always the case  in canonical gravity we consider only globally hyperbolic spacetimes that admit a $(3+1)$ split.

This article largely expands and gives complete proofs of some 
results that have been advanced in \cite{Pons:2009cd}.
In Section \ref{gaugefixing} we describe two alternative but related
procedures for constructing observables. These procedures start from
gauge-fixing conditions that explicitly depend on the coordinates in
such a way that the gauge choice is equivalent to a choice of
intrinsic coordinates. Both procedures amount to sending a point $p$
in the space of field configurations, representing a
solution of the field equations, to a point $p_{{}_{\!G}}$ where the
gauge conditions are fulfilled. An extended subsection \ref{active}
describes the first procedure in which observables are constructed
through active gauge transformations. This procedure delivers an
expression for the transformed field that can be solved explicitly
for a set of functions that determine the finite transformation.
These so-called ``descriptor'' functions depend on phase space
variables. Thus we are able to characterize all observables as
functional invariants in a generic manner that, to the best of our
knowledge, was not previously known . We find that every field -
including the lapse and the shift - has an associated observable, as
well as any functional combination of fields. Furthermore we are
able to state the equations of motion of the invariants - negating
claims in the literature that invariants are constants of motion.
The second procedure, described in \ref{trans}, amounts to
considering a passive coordinate transformation from $p$ to
$p_{{}_{\!G}}$. We establish the general relation between the active
canonical gauge transformations and the corresponding passive
coordinate transformation. This latter transformation is none other
than the transformation to intrinsic coordinates taking into account
the geometric character of each field.

We should mention here that our usage of ``passive" and ``active" transformations conforms with that in the community, described in e.g. \cite{Wald}, \cite{Rovelli:2004tv}. That is, passive diffeomorphism are always understood as coordinate transformations, whereas an active diffeomorphism is a mapping of a manifold to itself that induces pull-backs of the tensor-fields of the manifold. Although mathematically distinct - and we will make this distinction also in Appendix A - in many cases they can be made to be two sides of the same coin. The relation of passive and active transformations to each other, to dynamical symmetries of the Einstein field equations and to gauge transformations is treated in \cite{LP}.

In Section \ref{properties} we examine some properties of the
observables that can be derived from their definition, regardless of
their specific construction in a given theory. We show that they
exhibit a natural dependence on the time coordinate
 - the time coordinate  of the corresponding observer,
which, in the case that the observer sits at $p_G$, is the intrinsic time
and that their Hamiltonian dynamics
is in agreement with the dynamics of the gauge fixed fields at
$p_{{}_{\!G}}$. We also show that one can extract constants of
motion - and Noether generators - with no explicit time
dependence out of the invariants. We connect our findings with
the notion of ``evolving constants of motion''
\cite{Rov1,Rovelli:1989jn,Rovelli:1990jm}.

In \ref{secdirac} we give a simple geometric proof that the Poisson
bracket of the invariants associated with two given fields turns out
to be the invariant associated with the Dirac bracket of these
fields. This can be understood as giving a symmetry-based
interpretation of a proof by Thiemann\cite{thiemann06} based on a
formal series expansion. Furthermore our proof also includes the
lapse and the shift functions.

In \ref{limits} it is shown that the functional invariants can also
- rather intriguingly  - be understood as
limits of canonical maps. This provides an alternative route to the
results in section \ref{secdirac}.

Our main results are summarized in the concluding section, where we
interpret our findings in the light of related work on observables
in generally covariant theories.

We devote Appendix A to a more detailed consideration of those
points where, as a consequence of our belief in the central role of
diffeomorphisms in phase space, our ideas deviate from the opinions
of others in the canonical gravity community.  In the form of a
dialogue, we treat amongst others the issues of gauge
transformations and gauge generators, and the different roles of
gauge generators and the Hamiltonian. In  Appendix B  we prove a
lemma about appropriately redefined constraints.  This proof is a 
reelaboration of an earlier proof by Thiemann \cite{thiemann06}, in which he
investigated specific linear combinations of the secondary
first-class constraints with the property that the new constraints
have strongly vanishing Poisson brackets. This technical trick
significantly facilitated the explicit construction of  observables
as well as clarifying relations between Poisson and Dirac brackets
of observables.

\section{Constructing Observables through Coordinate Dependent Gauge Fixings}
\label{gaugefixing}

We recall \cite{pss1997pr} that the Legendre-projectable
infinitesimal passive coordinate transformations are of the form
\beq 
\tilde x^\mu = x^\mu -  [n^\mu(x) \xi^0(x; g_{ab}) - \delta^\mu_a
\xi^a(x; g_{ab}) ]\Delta s, \label{projtrans} 
\eeq  
 ($\Delta s$ is the infinitesimal parameter associated with the transformation.) This decomposition was
 employed by Bergmann and Komar in \cite{bk} in their endeavour to connect the diffeomorphism group of general relativity
with the algebra of constraints obtained in its canonical formulation, though they did not recognize that it followed from the requirement of projectability.
The generator of the corresponding active canonical phase space transformation is
\begin{equation}
   G_{{\xi }}(t) =    P_{\mu} \dot\xi^{\mu} + ( {\cal H}_{\mu}
+ N^{\rho} C^{\nu}_{\mu \rho} P_{\nu}) \xi^{\mu} . 
\label{thegen}
\end{equation}
(Repeated indices signify both a sum over the discrete index and a
3-dimensional integration over the spatial coordinates, unless
otherwise noted. Different indices correspond to different
coordinates.) In this expression $N^0 := N$ is the lapse, and the
$N^a$ are the shift. The $P_0$ and $P_a$ are their conjugate
momenta. They vanish as primary first class constraints. The normal
to the fixed coordinate time $t = constant$ hypersurface is $n^\mu =
\{N^{-1}, -N^{-1} N^a \}$. The ${\cal H}_\mu$ are the secondary
constraints that result from the preservation under time evolution
of the primary constraints $P_\mu \approx 0$. The $C^{\nu}_{\mu
\rho}$ are the structure coefficients in the algebra of the
${\cal H}_\mu$  under the Poisson bracket. The descriptors $\xi^{\mu}$ are arbitrary functions
of the spacetime coordinates as well as the fields other than the
lapse and shift. $\dot\xi^{\mu}$ is the time derivative of the
descriptors, which includes, in the case when $\xi^{\mu}$ depend
on fields, the implicit time dependence for
these fields as given by the dynamics. $G_{{\xi }}(t)$ acts at a single time $t$. In
order to produce the full infinitesimal action of an element of the
gauge group we need to specify the descriptors $\xi^\mu$ for all
values of the coordinate $t$. 

\subsection{Spacetime scalars as intrinsic coordinates}
\label{onlyscalars}

The coordinate dependent gauge fixing  program that we will
implement consists in using an appropriate set of four independent
scalar field functions $X^\mu$  in  a given spacetime and then
taking them as an``intrinsic" system of coordinates. This implies
that the program is only feasible for backgrounds admitting a scalar
coordinatization\footnote{That our real world admits such coordinatizations, at least in a 
spacetime region,  is shown in particular in \cite{Rovelli:2001my}.}. One possibility is to employ functions of Weyl scalars that are obtained from the Weyl conformal tensor\cite{berg-kom60}. This option is also in principle available in non-vaccum spacetimes with material field sources\cite{ps04}. However, one must be aware that in spacetimes with some Killing symmetries, it is likely that these scalars will not be independent and functions of them could  not then play the role of an intrinsic coordinatization. Let us review here a variation of the proof given
in \cite{ps04} that the intrinsic coordinate fields must be
spacetime scalars.  We interpret a choice of intrinsic coordinates
$X^\mu(x)$as a coordinate transformation from the coordinates
$x^\mu$ to $X^\mu$. Suppose that instead of starting with
coordinates $x^\mu$ we start instead with coordinates $f^\mu(x)$
before transforming to the intrinsic coordinate system $X_{f}(f(x))$.
Then the demand of invariance under the passage from $x^\mu$ to
$f^\mu(x)$ is the demand that the coordinate transformation from
$X^\mu$ to $X^\mu_f$ must be the identity transformation, i.e.,
invariance is precisely the demand that $X^\mu(x) = X_f^\mu(f(x))$.
This is the condition that $X^\mu(x)$ is a spacetime scalar.

The idea of using a set of four scalars can be traced back to
Einstein's hole argument that spacetime points can only be defined
and distinguished by values of physical fields or positions of
physical objects, \cite{stachel}, and has been stressed in
\cite{bergmann61}, \cite{dewitt62},\cite{ishamkuchar},
\cite{hartle91},\cite{Lusanna:2003im}. One either needs external reference objects like
dust \cite{brownkuchar} or GPS satellites \cite{Rovelli:2001my}, or
one identifies internal scalars, like in the Weyl-scalar program
initiated by Komar and Bergmann, \cite{berg-kom60},
\cite{bergmann61}.

\subsection{Constructing observables through active gauge transformations}
\label{active}

 The  gauge fixing conditions have the form \beq
\chi^\mu(x) := x^\mu - X^\mu(x) = 0. \label{gfconstr} \eeq Notice
that this gauge-fixing condition is explicitly coordinate dependent.
This coordinate dependence is mandatory and indeed, cf.
\cite{Sundermeyer:1982gv,ps95cqg,pss1997pr} one can formally prove
that this is the manner in which one guarantees that the resulting
dynamical evolution is never ``frozen''.

For most of our considerations, the arena will be the space ${\cal
S}$ of on-shell field configurations, i.e., fields obeying the
equations of motion.  This space is a subset of the much bigger space of 
general field configurations. An infinitesimal gauge transfomation acts
on this bigger space with the ordinary Poisson bracket, and its action 
can be restricted to ${\cal S}$ because the generators of gauge transformations 
define an action which is tangent to ${\cal S}$. This means that we do not need 
to know the off-shell extension of the on-shell field configurations for the action 
of the gauge generators to be well defined on ${\cal S}$. A point $p$ in ${\cal S}$ is in fact an entire spacetime 
with the fields - solution of the EOM - described in a particular coordinatization. 
For practical purposes, though, it will be enough to work in a coordinate patch of a 
given chart. To every point $p$ there is associated an "observer", or "user", 
who is using such a coordinatization to describe the fields in spacetime. 
In particular the time coordinate is a 
label for a foliation of the spacetime into spacelike hypersurfaces - at least in a region of it.  
The gauge generators, acting through the Poisson brackets, are used to construct finite gauge
transformations, realizing active diffeomorphism-induced
transformations at a fixed value of the spacetime coordinates. 
These gauge transformations define equivalence classes within ${\cal S}$, which
we call orbits of gauge equivalent spacetimes, or gauge orbits for
short. A whole gauge orbit represents a unique physical 
state\footnote{ Possible different understandings - and misunderstandings - of what a physical state 
is are dealt with in appendix A}, and its
different points correspond to different coordinatizations. One
can pass from one coordinatization to another by a passive
diffeomorphism. This gauge transformation is however not a dynamical
evolution (cf.  appendix A\footnote{ In appendix A we review the differences between
Bergmann's and Dirac's approaches to gauge transformations, and the consequences thereof. 
The incompleteness of Dirac's view in \cite{Dira64} is analyzed in
\cite{Pons:2004pp}})  because whereas a gauge 
transformation - different from the identity - maps 
a point $p$ into a different point $p'$, the dynamical evolution 
takes place entirely within every point $p$ in ${\cal S}$ - because every point 
represents a solution of the EOM.

Consider the point $p$ in ${\cal S}$ and let the fields at $p$ be
$X^\mu(x), \Phi^A(x)$, where $X^\mu(x)$ are the set of selected
scalars and $\Phi^A(x)$ denote all the remaining fields  or field
components. There should\footnote{If the chosen set of scalars
allows for a good coordinatization of the spacetime, no
Gribov ambiguitites can appear. Let us notice, though,
that our considerations are local, and that we can restrict
ourselves to a region of the spacetime where the scalar
coordinatization works well.} exist a finite gauge transformation
that moves this point $p$ in the gauge orbit to the unique point,
$p_{{}_{\!G}}$, that satisfies the gauge fixing constraints
(\ref{gfconstr}). We will also assume a trivial topology for the
orbit space - or at least in the region  in which we will work - so
that a diffeomorphism-induced transformation connected with the
identity will suffice. Therefore a finite gauge transformation
accomplishing our purpose has the form \cite{ps04}
$$ V_{\xi}(s,t) =  exp\left(s  \{ - , \,
G_{\xi }(t)\} \right) \,,
$$
with a given set of finite descriptors $\xi^\mu$. The parameter $s$ labels a trajectory within the gauge
orbit. By convention we will assume that at $s=1$ we reach the  point $p_{{}_{\!G}}$ in the orbit where the
gauge fixing constraints are satisfied. In particular, if we consider the scalars $X^\mu(x)$, we will have
\bea
X^\mu(x)& \rightarrow& {\hat X}^\mu (x,s) =  exp\left(s  \{ - , \,
G_{\xi }(t)\} \right) X^\mu(x) \nonumber \\
 &=&  X^\mu(x) + s  \{ X^\mu(x), G_{\xi }(t) \} \nonumber \\
 &+& \frac{s^2}{2}
 \left\{  \{ X^\mu(x), G_{\xi }(t) \}, G_{\xi }(t) \right\} + \cdots \,,
\eea
and the gauge fixing requirement is ${\hat X}^\mu (x,1) =: \hat X^\mu(x) = x^\mu$.
(Henceforth ``hatted' variables denote variables satisfying the gauge  fixing conditions.) This is an equation that
determines the descriptors $\xi(x)$, and we will obtain a unique solution for them in section \ref{descrip}.
Thus, to any point $p$ in some gauge orbit in ${\cal S}$ we associate a system of descriptors. Once the
descriptors are determined we can proceed to apply the gauge transformation to all the remaining fields,
\beq
{\hat \Phi}^A =exp\left(  \{ - , \,
G_{\xi }\} \right) {\Phi^A}  =:  {\cal F}_{\Phi^A} [X,\Phi;\xi]\,.
\label{preinv}
\eeq
\subsubsection{Solving for the descriptors}
\label{descrip}

Let us now solve for the descriptors $\xi^\mu$ required in
(\ref{preinv}). For this purpose it will be convenient to work with
the linear combination of secondary first class constraints that has
been  introduced by Henneaux and Teitelboim \cite{henneauxt94}
and further  exploited by Dittrich \cite{dittrich07} and Thiemann
\cite{thiemann06}. We set, at the fixed coordinate time $t$, 
\beq
{\cal A}^\mu_{\ \nu'} := \left\{ X^\mu, {\cal H}_{\nu'} \right\},
\label{calA} 
\eeq 
with inverse ${\cal B}^\alpha_{\ \beta'}$, i.e.,
${\cal B}^\alpha_{\ \beta'} {\cal A}^{\beta'}_{\ \nu''} =
\delta^\alpha_{\nu''}$.  (Here we introduce the convention that
primed indices represent evaluation at primed spatial coordinates,
and $\delta^\mu_{\nu'}:= \delta^\mu_\nu \delta^3 (x - x')$.)  We
define 
\beq \overline{\cal H}_\mu := {\cal B}^{\alpha'}_{\ \mu}
{\cal H}_{\alpha'}, 
\label{hbarra} 
\eeq and then rewrite $ \xi^{\mu}
{\cal H}_{\mu} = \overline{\xi}^{\mu} \overline{\cal H}_{\mu}$,
where $\overline{\xi}^\mu = {\cal A}^{\mu}_{\ \nu'} \xi^{\nu'}$.
Notice that therefore \beq \left\{ X^\mu, \overline{\cal H}_{\nu'}
\right\} \approx {\cal B}^{\alpha''}_{\ \nu'} {\cal A}^\mu_{\
\alpha''} = \delta^\mu_{\nu'}, \eeq where the weak equality
signifies that terms proportional to ${\cal H}_\mu$ have been
dropped, or in other words, we evaluate on the original first class
constrained hypersurface \footnote{An outcome of the construction is
that $\{\overline{\cal H}_\mu\,,\overline{\cal H}_\nu\}$ is strongly
vanishing, instead of weakly vanishing. This result is
 derived  in \cite{thiemann06}. We give an alternative symmetry-based proof
in appendix B.}.  Of course the change from the original ${\cal
H}_\mu$ to their linear combination $\overline{\cal H}_\nu$
according to (\ref{hbarra}) is only possible locally.

 We will assume that the lapse and shift
are not involved in the construction of the scalar fields
$X^\mu(x)$. In this case only the ${\cal H}$ contribution to the
generator $G_\xi$ in (\ref{thegen}) is relevant. 
Thus the gauge transformed scalar fields, transformed to
 $p_{{}_{\!G}}$, are \beq \hat X^\mu(x) =
x^\mu = exp\left(  \{ - , \, \overline{\xi}^{\nu'} \overline{\cal
H}_{\nu'}\} \right) X^\mu(x)
 \approx  X^\mu(x) +  \overline{\xi}^\mu,
\eeq and we can therefore solve on shell for \beq
\overline{\xi}^\mu[X(x);\,x] = x^\mu - X^\mu(x) =: \chi^\mu(x)\,,
\label{thedescrip} \eeq where $\chi^\mu$ are the gauge fixing
constraints introduced in (\ref{gfconstr}).

Although we have obtained a simple closed-form for functionals
associated with the descriptors in the basis $\overline{\cal
H}_{\nu}$ for the Hamiltonian constraints, the construction of
$\overline{\cal H}_{\nu}$ can of course be difficult in practice due
to the need to invert the matrix ${\cal A}^\mu_{\ \nu'}$. But we
will nevertheless be able to prove some interesting formal results
in section \ref{properties}.

\subsubsection{The observables  associated with fields other than the lapse and shift}
 \label{theinvariants}

We are now in the position to derive an expression for the
observables  in terms of the gauge fixing
conditions.  It is methodologically convenient to first consider the observables 
associated with fields other than the lapse and shift. This means that  the on shell action of the 
gauge generator (\ref{thegen}) is given just by ${\cal H}_{\mu}\xi^{\mu}$. 
Thus throughout this section, the fields $\Phi^A$ do not include the lapse and shift. In subsection \ref{anyfield} this restriction is lifted.
We keep working with the basis $\overline{\cal H}_{\nu}$
for the Hamiltonian constraints, and to make the following
considerations easier to follow we temporarily attach a subscript
$p$ to the arguments of the functionals in (\ref{preinv}),
signifying that they refer to the point $p$ in the gauge orbit, \beq
{\hat \Phi}^A =exp\left(  \{ - , \, G_{\overline{\xi}_p }\} \right)
{\Phi^A_p}  =: {\cal F}_{\Phi^A} [X_p,\Phi_p;\overline{\xi}_p]\,,
\label{hatphi} \eeq where $\overline{\xi}_p$ are taken as functions
of the spacetime coordinates only, whose determination at $p$ is
given by $\chi^\mu_p= x^\mu - X^\mu_p(x)$. Of course, had we started
with another point $p'$ in the gauge orbit, we would have written
$$
{\hat \Phi}^A  =  {\cal F}_{\Phi^A} [X_{p'},\Phi_{p'};\overline{\xi}_{p'}]\,,
$$
with the {\sl same} functional form, because it is the same gauge
transformation, see (\ref{preinv}), with another set of descriptors.
So we have \beq {\cal F}_{\Phi^A}
[X_p,\Phi_p;\overline{\xi}_p]={\cal F}_{\Phi^A}
[X_{p'},\Phi_{p'};\overline{\xi}_{p'}]\,. \label{preinv2} \eeq
Notice that, since they are determined by the field configurations
$X_p$, the descriptors used to send these field configurations at
$p$ to their expressions ${\hat
X}=:X_{p_{{}_{\!G}}},\,{\hat\Phi}=:\Phi_{p_{{}_{\!G}}}$, at
$p_{{}_{\!G}}$ are functionals of $X_p$. One can then write
\footnote{We assume - and it will prove crucial for the procedure to
succeed - that the functionals $\xi^\mu$ may carry explicit
dependencies on the spacetime coordinates $x^\mu$. Remember that
when we move from point to point, $p\to p'$ in the gauge orbit
through an active diffeomorphism-induced transformation, the
spacetime coordinates do not change.}, generically,
$\overline{\xi}^\mu_p(x) = \chi^\mu_p(x)= x^\mu - X^\mu_p(x)$, and
define the new functionals \beq {\cal I}_{\Phi^A}
[X_{p},\Phi_{p};\,x]:={\cal F}_{\Phi^A}
[X_{p},\Phi_{p};\overline{\xi}_{p}^\mu]_{\big|_{\overline{\xi}_{p}=
\chi_{p}}} \,. \label{xi} \eeq

It is  important to understand  that
the {\sl same} functionals $\chi^\mu$ work for any point $p$,
because $p$ is a generic point in the gauge orbit. That is, for
another point $p'$, we will have $\overline{\xi}^\mu_{p'}(x)=
\chi^\mu_{p'}(x)= x^\mu - X_{p'}^\mu(x)$. Thus, using
(\ref{preinv2}), 
\bea 
{\cal I}_{\Phi^A}[X_p,\Phi_p;\,x]&=&{\cal
F}_{\Phi^A}[X_{p},\Phi_{p};\overline{\xi}_{p}^\mu]_{\big|_{\overline{\xi}_{p}
= \chi_{p}}}  \nonumber \\
&=& {\cal F}_{\Phi^A} [X_{p'},\Phi_{p'};\overline{\xi}_{p'}^\mu]_{\big|_{\overline{\xi}_{p'}= \chi_{p'}}}  \nonumber \\
&=&
{\cal I}_{\Phi^A}[X_{p'},\Phi_{p'};\,x]\,.
\label{invfunct}
\eea

Equation (\ref{invfunct}) expresses the invariance of the
functionals ${\cal I}_{\Phi^A}$. These functionals are observables.
In terms of infinitesimal transformations the invariance
(\ref{invfunct}) reads \beq \{ {\cal I}_{\Phi^A}, G_{\eta }\}\approx
0\,, \label{theinv} \eeq for arbitrary descriptors $\eta$ in $G$.
Due to this arbitrariness and the generic form (\ref{thegen}),
(\ref{theinv}) is equivalent to $$\{ {\cal I}_{\Phi^A}, {\cal
H}_\mu\} \approx 0, \,\,\,\,\,\,\,\,\,\,\,\,\,\, \{{\cal
I}_{\Phi^A}, {\cal P}_\mu\} \approx 0,$$ and these are the defining
conditions for observables.  Observe that since $\{ -,\, G_{\eta
}\}$ is tangent to the gauge orbit, the variations of the fields in
the functional ${\cal I}_{\Phi^A}$ in (\ref{theinv}) are always
along the gauge orbit, and thus we need only information of the
functionals on shell to be able to compute (\ref{theinv}).

A subtlety not to be overlooked in the definition (\ref{xi}) is the
following: the substitution of the descriptors $\overline{\xi}_{p}$
by the gauge fixing constraints (which do not vanish in $p\neq
p_{{}_{\!G}}$) is made {\sl after} the functional ${\cal
F}_{\Phi^A}$ has been computed with a descriptor that has no
dependence on the fields; or to say it in another way, the
descriptors used in (\ref{hatphi}) have vanishing Poisson brackets
with all the fields. More on this will be said in section
\ref{limits}.

At this point, some further comments are in order. First, it is
worth noticing that the invariants ${\cal I}_{\Phi^A}$ will in
general be non-local as regards the spatial coordinates, due to the
nesting of commutators in the expansion of the functionals in terms
of the fields and their space derivatives; we 
encounter here a result first obtained by Torre in
\cite{Torre:1993fq}. Second,  citing in advance a result from the following section,  
there is an invariant associated with any
field, including lapse and shift. Third, the method above can also
be used to define invariants associated with any functional of the
fields. And fourth, the observables ${\cal
I}_{\Phi^A}$ can be interpreted in two equivalent ways. On one hand,
an observer with an on-shell field configuration
 at $p$ has a prescription $\Phi^A\to {\cal I}_{\Phi^A}$ for
associating an invariant with any field, and she knows that the
description of her solution provided by the
invariants will coincide with that of any other observer that uses
the {\sl same recipe} to obtain the invariants, but she always
remains  at $p$. On the other hand, if such an observer decides to
use these functionals in order to work with the new fields ${\hat
\Phi}^A :={\cal I}_{\Phi^A}[X_p,\Phi_p;\,x]$, then this means that
she has been able to obtain, 
with the redefiniton of the fields  ${\Phi^A}\to{\hat\Phi}^A$ the 
description of the observer at $p_{{}_{\!G}}$, just reflecting  
the active view of diffeomorphism transformations. Notice that as long as she decides to work 
with the new fields ${\hat\Phi}^A$ - the observables - as the fields of her spacetime, 
everything in her new description is as if her original coordinates played the role of the 
intrisinc coordinates. We comment on the complementary passive view in subsection \ref{trans}.

\vspace{4mm}

With the explicit solution (\ref{thedescrip}) for the descriptors,
the expansion of the invariant ${\cal I}_{\Phi^A}$ in (\ref{xi}) becomes
\bea
{\cal I}_{\Phi^A} &\approx& exp\left(  \{ - , \,
\overline{\xi}^{\nu} \overline{\cal H}_{\nu}\} \right)\Phi^A{}_{\big|_{\overline\xi= \chi}} = \Phi^A
+    \chi^\mu\{ \Phi^A , \,\overline{\cal H}_{\mu}\} \nonumber\\
&+& \frac{1}{2!}\chi^\mu\chi^\nu\{\{ \Phi^A , \,
 \overline{\cal H}_{\mu}\}, \,\overline{\cal H}_{\nu}\} \nonumber\\
 &+&
  \frac{1}{3!}\chi^\mu\chi^\nu\chi^\rho\{\{\{ \Phi^A , \,
 \overline{\cal H}_{\mu}\}, \,\overline{\cal H}_{\nu}\}, \,\overline{\cal H}_{\rho}\}+\ldots\nonumber\\
&=:&\sum_{n=0}^{\infty} \frac{1}{n!}\chi^n \{ \Phi ,
\,\overline{\cal H}}\}_{(n)\,. \label{expansion} \eea (In the last
line we have adopted a simplifying notation where indices in $\chi$
saturate with indices of $\overline{\cal H}$ and $\{ \Phi ,
\,\overline{\cal H}\}_{(n)}$ is interpreted as the repeated nesting
of $n$ Poisson brackets with $\overline{\cal H}$ in the right hand
side). With different notation, this expression appeared in the
literature in \cite{thiemann06} as his equation (2.8)and in
\cite{dittrich07}  as her equation (5.23). Here we have arrived at
(\ref{expansion}) by a symmetry-inspired procedure, as the effect of
the finite gauge transformation that sends $p$ to $p_{{}_{\!G}}$.
This specific gauge transformation is determined once the set of
scalar fields associated with the gauge fixing has been selected. An
advantage of the present formulation is that one can send all the
fields from $p$ to $p_{{}_{\!G}}$, and this includes the lapse and
shift. In this general case one must use the full gauge generator
(\ref{thegen}) and it is worked out in section \ref{anyfield}.

The gauge invariance of (\ref{expansion}) is guaranteed by the
construction procedure, as long as the series expansion is
convergent, which is expected at least in a neigborhood of
$p_{{}_{\!G}}$. Note notwithstanding that one can directly verify
the gauge invariance of ${\cal I}_{\Phi^A}$ by checking the
vanishing on shell of $\{ {\cal I}_{\Phi^A},\, \overline{\cal
H}_\mu\}$ with use of the on shell expansion (\ref{expansion}). It
is crucial in this respect, as noticed  already  in
\cite{thiemann06}, that the Poisson
brackets of the constraints $\overline{\cal H}_\mu$ among themselves
are quadratic in the constraints (see Appendix B). One can proceed as
follows. Let us define
$$
B_{{}_\Phi}^{(n)} := \chi^n \{ \Phi , \,
 \overline{\cal H}\}_{(n)}\,,
 $$
with $B_{{}_\Phi}^{(0)}= \Phi$. Then 
\beq 
\{B_{{}_\Phi}^{(n)},\,\overline{\cal H}\} \approx - n
B_{{}_{\{\Phi,\,\overline{\cal H}\}}}^{(n-1)} +
B_{{}_{\{\Phi,\,\overline{\cal H}\}}}^{(n)}\,, \eeq from which, \bea
&&\{ {\cal I}_{\Phi}, \overline{\cal H}_\mu\}= \{\sum_{n=0}^{\infty}
\frac{1}{n!}B_{{}_\Phi}^{(n)} ,\, \overline{\cal H}_\mu\}
\nonumber\\ &\approx& - \sum_{n=1}^{\infty} \frac{1}{(n-1)!}
B_{{}_{\{\Phi,\,\overline{\cal H}\}}}^{(n-1)} + \sum_{n=0}^{\infty}
\frac{1}{n!} B_{{}_{\{\Phi,\,\overline{\cal H}\}}}^{(n)}=0\,. 
\eea

\vspace{4mm}

It is worth noticing that the proof of invariance given above does not depend of the fact that the gauge
fixing constraints $\chi^\mu$ are made up with scalar fields $X^\mu$. If, instead of using the scalar fields
$X^\mu$ in the process to define the basis $\overline{\cal H}_\mu$ for the Hamiltonian constraints, one
uses another set of fields - or field components -, the proof of invariance remains intact.

On the other hand, if non-scalars were used for the gauge fixing, it
is very likely that Gribov ambiguities will appear.
Suppose for instance that the gauge fixing were implemented with a
vector $J^\mu(x)$, so one should make a change of coordinates
$x\to\hat x$ such that $\hat x^\mu-{\hat J}^\mu(\hat x)=0$.
Considering the rules to transform a vector under diffeomorphisms:
$J^\mu(x)\to \hat J^\nu(\hat x) =J^\mu(x)\frac{\partial \hat
x^\nu}{\partial x^\mu }$, one should look for a transformation
realizing ${\hat J}^\nu(\hat x) = \hat x^\nu$.

Since the intrinsic vector field is $\vec J = J^\mu(x)\frac{\partial }{\partial x^\mu }$,
the equation to obtain the intrinsic coordinates is nothing but $\vec J\,\hat x^\nu(x)
= \hat x^\nu(x)$. Thus we look for four  eigenfunctions of $\vec J$ with unit eigenvalue.

If we have four such  eigenfunctions 
$f^\mu(x),\ \vec J\, f^\nu = f^\nu$, any linear numerical
matrix $A$ will introduce an ambiguity $f^\mu\to \tilde f^\mu = A^\mu_\nu f^\nu$.

\vspace{4mm}

As argued at the beginning of section \ref{onlyscalars}, a proper
gauge fixing needs to be performed with spacetime scalars, and we
will maintain this requirement throughout.

\subsubsection{Observables associated with  fields including lapse and shift}
\label{anyfield}

Here we extend the results of the previous section to include the
observables associated with the lapse and shift fields.

Recall that the gauge generator (\ref{thegen}) is,
$$
   G_{{\xi }}(t) =    P_\mu \dot\xi^\mu + ( {\cal H}_\mu
+ N^{\rho} C^{\nu}_{\mu\rho} P_{\nu}) \xi^\mu\,,
$$
where the  descriptors $\xi^\mu$ are arbitrary functions which may depend
of course on the coordinates but also on the fields other than lapse and
shift. $\dot\xi^\mu$ is read as
$$\dot\xi^\mu = \frac{d}{d\,t}\xi^\mu = \frac{\partial}{\partial t} \xi^\mu
+ N^\rho \{\xi^\mu\,,{\cal H}_\rho\}\,,
$$
so that the explicit dependence on the time parameter is accounted for in the first term whereas in the
second term the implicit time dependence through the fields is reflected through their own dynamics.

When all fields are considered, one must observe that the gauge fixing
constraints $\chi^\mu= x^\mu - X^\mu =: \chi^{(1) \mu}
 $ have secondary descendants:
$$
\frac{d}{d\,t}\chi^\mu = \delta^\mu_0 -  N^\rho \{X^\mu\,,{\cal H}_\rho\} \,,
$$
and thus the lapse and shift become determined by the secondary
gauge fixing constraints \beq \chi^{(2) \mu} :=\delta^\mu_0 - {\cal
A}^\mu_{\ \rho}N^\rho\approx 0\,, \label{secondgf} \eeq Preservation
of these constraints in time leads to the determination of the
arbitrary functions in the Dirac Hamiltonian and the gauge is completely fixed. Note that, by
definition, it is only at $p_{{}_{\!G}}$ that these gauge fixing
constraints are satisfied.

Now we follow the same steps taken in \ref{descrip}, but with the number of constraints doubled.
Our $8$ gauge fixing constraints  $ \chi^{(i) \mu} = (\chi^\mu,\, \dot\chi^\mu )$ can be used
to change the basis
of the $8$ first class constraints, $\zeta_{(j) \nu}= ({\cal H}_\nu,\,P_\nu )$ to another basis
${\overline \zeta}_{(j) \nu}=(\overline{{\overline {\cal H}}}_\nu,\,{\overline P}_\nu) $ so that
$\{\chi^{(i) \mu},\,{\overline \zeta}_{(j) \nu} \} \approx - \delta^i_j \delta^\mu_\nu$. These new
${\overline \zeta}_{(i) \mu}$ will have strongly (instead of weakly) vanishing Poisson brackets among
themselves at any point $p$ in the gauge orbit. This setting is  convenient because it makes
possible an easy determination of the descriptors associated with the specific gauge transformation
that sends the field configurations at $p$ to their corresponding fields at $p_{{}_{\!G}}$.

The matrix of the gauge fixing constraints with the first class constraints,
$$\{\chi^{(i) \mu},\,\zeta_{(j) \nu}\} = \left(\begin{array}{cc}- A^\mu_{\ \nu} & 0
 \\ - \{{\cal A}^\mu_{\ \lambda},\,{\cal H}_\nu\} N^\lambda
 & - {\cal A}^\mu_{\ \nu} \end{array}\right)\,,
$$
has the inverse
\bea
{\cal M}^{i \mu}{}_{\ j \nu} &:=&(\{\chi^{(i) \mu},\,\zeta_{j \nu}\})^{-1} \nonumber\\ &=& \left(\begin{array}{cc}- B^\mu_{\ \nu} & 0
  \\ B^\mu_{\ \lambda }B^\rho_{\ \nu }N^\sigma\{A^\lambda_{\ \sigma }, {\cal H}_\rho\} 
& - {\cal B}^\nu_{\ \nu}
\end{array}\right)\,,\nonumber
\eea
(where ${\cal B}^\mu_{\ \rho}$ was defined before, see \ref{descrip}, as the inverse matrix of
${\cal A}^\mu_{\ \rho}:=\{X^\mu\,,{\cal H}_\rho\}$) and defines the new basis of first class constraints as ${\overline \zeta}_{(j) \nu}
=- {\cal M}^{i \mu}{}_{\ j \nu} \zeta_{i \mu}$.
We obtain
$$
{\overline P}_\mu = {\cal B}^\rho_{\ \mu}P_\rho\,,
$$
and
$$
\overline{{\overline {\cal H}}}_\nu
= {\cal B}^\rho_{\ \nu}\left( {\cal H}_\rho - {\cal B}^\mu_{\ \lambda}N^\sigma \{A^\lambda_{\sigma},\,{\cal H}_\rho\} P_\mu \right)\,. \label{calh0}
$$
As a consequence, to express the gauge generator in the new basis we need to implement
$$
P_\mu= {\cal A}^\rho_{\ \mu}{\overline P}_\rho\,, \ \ {\cal H}_\rho= {\cal A}^\nu_{\ \rho}\overline{{\overline {\cal H}}}_\nu
+ {\cal B}^\mu_{\ \lambda}N^\sigma \{A^\lambda_{\ \sigma},\,{\cal H}_\rho\}P_\mu\,.
$$
With these substitutions, the gauge generator (\ref{thegen}) becomes
\bea
   G_{{\xi }}(t)&=& {\cal A}^\nu_{\ \mu}{\overline P}_\nu \dot\xi^\mu + \left( {\cal A}^\nu_{\ 
\mu}\overline{{\overline {\cal
H}}}_\nu+ {\cal B}^\lambda_{\ \rho}N^\sigma \{A^\rho_{\ \sigma},\,{\cal H}_\mu\}P_\lambda \right. \nonumber\\
&+& \left. N^\sigma C^\lambda_{\mu\sigma}P_\lambda \right)\xi^\mu\,.
\eea
 Now we may consider the special case when $\xi^\sigma$ is such
that ${\overline \xi}^\nu:={\cal A}^\nu_{\ \sigma}\xi^\sigma$ is
field independent. We have
$$\dot\xi^\mu =
 \frac{d}{d\,t}({\cal B}^\mu_{\ \sigma}{\overline\xi}^\sigma)=\{{\cal B}^\mu_{\ \sigma},\,
 N^\lambda {\cal H}_\lambda\} {\overline\xi}^\sigma +
 {\cal B}^\mu_{\ \sigma} \dot{\overline\xi^\sigma}\,,
$$
and the gauge generator becomes
\beq
   G_{\overline{\xi }}(t) = {\overline P}_\nu\dot{\overline\xi^\nu} +\overline{ {\overline {\cal
H}}}_\nu{\overline\xi^\nu}
+ P_\mu N^\sigma{\cal S}^\mu_{\rho\sigma}{\overline\xi^\rho}\,,
\eeq
where ${\cal S}^\mu_{\rho\sigma}$ is defined as
\beq
{\cal S}^\mu_{\rho\sigma}= \{{\cal B}^\mu_{\ \rho},\,{\cal H}_\sigma\} + {\cal B}^\nu_{\ \rho}{\cal B}^\mu_{\ \gamma}\{{\cal A}^\gamma_{\ \sigma},\,{\cal H}_\nu\} + {\cal B}^\nu_{\ \rho}C^\mu_{\nu\sigma}\,.
\eeq
But ${\cal S}^\mu_{\rho\sigma}$ is just a constraint, in fact a linear combination of the Hamiltonian constraints. To see this let us use the fact that the matrices ${\cal A}$, ${\cal B}$, (with discrete and continuous indices as well) are inverses to each other. We obtain
$$
 {\cal A}^\beta_{\ \mu}{\cal A}^\rho_{\ \alpha}{\cal S}^\mu_{\rho\sigma} = -\{{\cal A}^\beta_{\ \alpha},\,{\cal
H}_\sigma\}
+ \{{\cal A}^\beta_{\ \sigma},\,{\cal H}_\alpha\} + C^\mu_{\alpha\sigma}{\cal A}^\beta_{\ \mu}\,,
$$
which, using the definition (\ref{calA}) of ${\cal A}^\beta_{\ \mu}$, becomes
\bea
 {\cal A}^\beta_{\ \mu}{\cal A}^\rho_{\ \alpha}{\cal S}^\mu_{\rho\sigma} &=& -\{X^\beta,\,\{{\cal H}_\alpha,\,{\cal
H}_\sigma\}\} + C^\mu_{\alpha\sigma}\{X^\beta,\,{\cal H}_\mu\} \nonumber\\
&=& - \{X^\beta,\,C^\mu_{\alpha\sigma}\}{\cal H}_\mu\,,
\eea
which proves our assertion. The gauge generator in the new basis is therefore
$ G_{\overline{\xi }}(t) ={\overline P}_\nu\dot{\overline\xi^\nu} + \overline{{\overline {\cal
H}}}_\nu{\overline\xi^\nu} + {\cal O}(2)\,,
$
where by ${\cal O}(2)$ we mean terms that are quadratic in the constraints. Since we always work on
shell, this last term is irrelevant and we discard it, obtaining a very simple expression for the
gauge generator in the new basis,
\beq G_{\overline{\xi }}(t) ={\overline P}_\nu\dot{\overline\xi^\nu} + \overline{{\overline {\cal
H}}}_\nu{\overline\xi^\nu} \,.
\label{gengen}\eeq

Finally, imposing the usual condition: $exp(\{-,\,G_{\overline{\xi}}\}) X^\mu =\hat X^\mu =  x^\mu$, we can
determine the descriptors that must be employed to transform from $p$ to $p_{{}_{\!G}}$. Their functional form on
shell is
$${\overline \xi}^\sigma \rightarrow
\chi^\sigma= x^\sigma - X^\sigma\,,\ \ {\dot{\overline\xi^\sigma}}\rightarrow{\dot{\chi^\sigma}}
= \delta^\sigma_0-N^\nu A^\sigma_{\ \nu},.$$

Thus for any field, including lapse and shift, the invariants are defined as in section
\ref{theinvariants} but with
the full gauge generator $G_{\overline{\xi }}$ given in (\ref{gengen}) and with the substitutions for the
descriptors  implemented after the action of the gauge generator, as prescribed above.

\vspace{4mm}

\subsection{Passive coordinate transformation from $p$ to $p_{{}_{\!G}}$}
\label{trans}

There must exist passive coordinate transformations that
correspond to the active transformation to the gauge fixed point
$p_{{}_{\!G}}$ on the gauge orbit, and we can easily find them.  Let us suppose that this passive
transformation takes the functional form $\hat x^\mu = f^\mu(x)$.
Since by assumption $X^\mu(x)$ transforms as a spacetime scalar
under this transformation, we have $\hat X^\mu (\hat x) = X^\mu
(x)$, so that the transformed fields at the original coordinate
location $x^\mu$ are $\hat X^\mu ( x) = x^\mu = X^\mu\left(
f^{-1}(x)\right)$. Equivalently 
\beq f^\mu(x) = X^\mu(x).
\eeq 
In
other words, and this is one of our key observations, 
the performance of the gauge transformation  - active view - to the solutions
satisfying the gauge conditions is equivalent to the performance of
a coordinate transformation  - passive view - from the original coordinate 
system to
intrinsic coordinates. {\it Every} dynamical field evaluated in this
intrinsic coordinate system will therefore be an invariant under
diffeomorphism-induced gauge transformations, and this includes the
lapse and shift.

Considering in particular the case of a  scalar field $\Psi$, as
described by the user sitting at $p$, the prescription dictated by
the passive coordinate transformation is $\hat \Psi(x) =
\Psi(X^{-1}(x))$ or, equivalently, $\Psi_{p_{{}_{\!G}}}(x) =
\Psi_p(X^{-1}(x))$. In the case of other, non-scalar, fields, they
transform according to their geometric properties. These observables
may be displayed as power series in the  coordinates
$x^\mu$ in the following manner. Repeated derivatives of the
identities $X^{-1}{}^\mu(X(x)) = x^\mu$ followed by substitution of
the Hamiltonian equations of motion will yield a Taylor expansion in
$x^\mu$ with coefficients $\frac{\partial^{k_0+\cdots + k_3}
X^{-1}}{\partial (x^0)^{k_0} \cdots  \partial (x^3)^{k_3} }$
evaluated at $x^\mu = X^\mu$.  The resulting expansions can be
obtained in a more efficient manner through active transformations,
as we now show.

\section{Properties of the observables}
\label{properties}
\subsection{Dynamics of the observables}
\label{dynprop}
\subsubsection{The equations of motion for the invariants  ${\cal I}_{\Phi}$}

The explicit dependence on time - through the determination of the
descriptors (\ref{thedescrip}) - makes the observables
(\ref{expansion}) time dependent; the implicit dependence is
canceled due to the invariance $\{{\cal I}_{\Phi},\,{\cal
H}\}\approx 0$. Considering fields $\Phi$ other than lapse and shift
(for which the following considerations can be extended
appropriately), 
\bea \frac{d}{d\,t} {\cal I}_{\Phi} &=&
\frac{\partial }{\partial \,t}{\cal I}_{\Phi} + \{{\cal
I}_{\Phi},\,N^\mu{\cal H_\mu}\}
\approx \frac{\partial }{\partial \,t}{\cal I}_{\Phi} \nonumber \\
&\approx& \{ \Phi , \,
 \overline{\cal H}_{0}\}
+\frac{1}{2}\left(\chi^0\chi^\nu\{\{ \Phi , \,
 \overline{\cal H}_{0}\}, \,\overline{\cal H}_{\nu}\} \right. \nonumber\\
&+& \left.
 \chi^\mu\chi^0\{\{
\Phi, \,
 \overline{\cal H}_{\mu}\}, \,\overline{\cal H}_{0}\} \right) +\cdots \nonumber \\
&\approx& \sum_{n=0}^{\infty} \frac{1}{n!}\,\chi^n \{\{ \Phi , \,
 \overline{\cal H}_0\}, \,\overline{\cal H}\}_{(n)} ={\cal I}_{\{ \Phi , \,
 \overline{\cal H}_0\}}\,,
\eea
where in the second line we have used
the strong vanishing of the Poisson brackets of the $\bar{\cal H}$,
cf. Appendix B. Thus the equations of motion for the invariants are
\beq\frac{d}{d\,t} {\cal I}_{\Phi}\approx{\cal I}_{\{ \Phi ,
\,\overline{\cal H}_0\}}\,. \label{dynob}\eeq The appearance of
$\overline{\cal H}_0$ in (\ref{dynob}) might come as a surprise, but
it is exacty what is needed in order for the fields at
$p_{{}_{\!G}}$ to satisfy the equations of motion. In fact, starting
at any point $p$ in ${\cal S}$, the invariants produce the
corresponding fields at $p_{{}_{\!G}}$: ${\cal
I}_{\Phi}[X_{p},\Phi_{p};\,x] = \Phi_{p_{{}_{\!G}}}$, and ${\cal
I}_{\{ \Phi , \,\overline{\cal H}_0\}}[X_{p},\Phi_{p};\,x]= {\{ \Phi
, \,\overline{\cal H}_0\}}_{p_{{}_{\!G}}}$. On the other hand, from
(\ref{hbarra}) we know that $\overline{\cal H}_0 = {\cal B}^{\mu}_{\
0} {\cal H}_{\mu}$. Also, at $p_{{}_{\!G}}$, the dynamical
stabilization of the gauge fixing constraints introduces the new
constraints (\ref{secondgf}),which may be written as \beq N^\mu -
{\cal B}^{\mu}_{\ 0} \approx 0\,, \label{nmu} \eeq thus fixing the
values of lapse and shift in terms of other fields. Note then that,
when computed at $p_{{}_{\!G}}$, $\overline{\cal H}_0 \equiv N^\mu
{\cal H}_\mu$, where we use Dirac's strong equality symbol, defined
in Appendix B. Thus, the content of (\ref{dynob}) is just \beq
\frac{d}{d\,t} \Phi_{p_{{}_{\!G}}} \approx \{ \Phi , \,N^\mu {\cal
H}_\mu\}_{p_{{}_{\!G}}}\,, \label{phieom}\eeq which is nothing but
the equations of motion for the fields - other than lapse and
shift - at $p_{{}_{\!G}}$ ! This result can be extended for the
lapse and shift fields by using techniques described in section
\ref{anyfield}.

\vspace{4mm}

This fact that observables are time dependent, already stressed in \cite{ps04},
contradicts claims of standard lore
asserting that observables are compelled to be constants of motion because
they have vanishing Poisson bracket with the generator of time evolution. 
The simple mistake in this claim is that it doesn't take into account the compulsory explicit time
dependence\footnote{ One can recall here the case of Galilean mechanics, 
were the boost generators are constants of motion with explicit time dependence, 
and therefore their Poisson bracket with the Hamiltonian is non vanishing.}
that is needed for a correct gauge fixing in generally covariant theories,
which has been proven in \cite{Sundermeyer:1982gv,ps95cqg}. Deeply connected with this mistake is the
confusion between gauge symmetry and dynamical evolution, also common in the literature and to which we
devote the Appendix A.

\subsubsection{From observables to ``evolving constants of motion"}

We continue to work for simplicity with fields other than the lapse and shift. We
shall show now how constants of motion - and Noether generators -
without explicit time dependence can be easily recovered form our
program.

Let us stress again the fact that the time dependence in the
invariants ${\cal I}_{\Phi}$, (\ref{expansion}), has two sources.
One is the explicit time dependence in $\chi^0 = t - X^0$, and the
other is the implicit time dependence in the on shell field
configuration (including the scalars $X^\mu$). But we have already
seen at the beginning of the previous section that this implicit
time dependence is non-existent because of the gauge invariance. In
fact, it is easy to separate these two dependencies and witness the
difference. For instance one could write ${\cal I}_{\Phi}$ with an
``explicit" time  $t_1$ and an ``implicit" 
 - i.e., the time argument in the field configuration - 
time $t_2$ and eventually 
realize\footnote{Note that for each point $p$ in $\cal S$, and given 
$t_1,\,t_2$, there is a point $p'$ where $\Phi_p(t_2,\vec x) 
= \Phi_{p'}(t_1,\vec x)$, for any field or field component $\Phi$.} 
 that
the only true dependence is in $t_1$. This result suggests that we
can consider the invariants as a one-parameter family of functionals
in phase space, which can be evaluated on on shell field
configurations at an arbitrary time $t_2$. Notice that as long as we
keep $t_1$ fixed, the invariants do indeed yield
 constants of motion.

Up to now our arena regarding the distinction between  ``explicit'' 
and ``implicit'' time dependence has been the space ${\cal S}$ of on shell
field configurations. Now, continuing with our previous construction
we move to phase space, where the variables are the fields
exhibiting only spatial dependence on the coordinates.  Let us
examine the invariants ${\cal I}_{\Phi}$ in phase space and its
remaining dependence on the ``explicit" time. In, fact, using
(\ref{dynob}) and differentiating repeatedly with respect to the
explicit time, we obtain \beq\frac{\partial^n}{\partial\,t^n} {\cal
I}_{\Phi}\approx{\cal I}_{\{ \Phi , \,\overline{\cal
H}_0\}_{(n)}}\,. \label{dynob2}\eeq where $\{ \Phi ,
\,\overline{\cal H}_0\}_{(n)}$ has the usual sense of nested Poisson
brackets.
 From (\ref{dynob2}) we can build the Taylor expansion in the $t$ parameter
\beq 
{\cal I}_{\Phi}\approx \sum_{n=0}^{\infty} \frac{t^n}{n!}\,
{\cal I}_{\{ \Phi , \,\overline{\cal H}_0\}_{(n)}}{}_{\big|_{t=
0}}\,. \label{evolv} 
\eeq 
(Recall that we think of these quantities
in phase space, i.e., the  invariants are
now dependent on the phase space variables, these phase space
variables being fields depending on the spatial coordinates.) Notice
that the coefficients of this expansion in the time parameter are
constants of motion, because they are invariants; since they have
vanishing Poisson brackets with the Hamiltonians they have no
explicit time dependence. In fact, as pointed out  before, fixing
the  explicit time parameter at any arbitrary value, the invariants ${\cal
I}_{\Phi}$ become constants of motion.

One may wonder what is then the role of this explicit time
dependence in the invariants? The answer has been given in equation
(\ref{phieom}), which is a consequence of (\ref{dynob}). It is
remarkable that the explicit time dependence in (\ref{evolv}) allows
us to fully recover the dynamics for the configuration of the fields
that satisfy the gauge fixing. Thus, two observers using the same
recipe to construct the invariants will be able to describe with the
help of these invariants the same physics {\sl at any time} - which
is their own original time coordinate  - because the
dynamical evolution is already built into the invariants 
 thanks to the explicit time dependence. With their
invariants, they will describe physical
evolution in an invariant way, through the explicit time dependence.
 Notice that from the point of view of a typical observer, sitting at 
some point $p$ in ${\cal S}$, the explicit time parameter appearing 
in his/her observables ${\cal I}_\Phi$ is just his/her original time parameter, 
that is, a labeling for the spacelike hypersurfaces foliating the 
spacetime - or at least a region of it -, satisfying the rather mild 
requirement of being an increasing function from past to future.
 In addition, for the observer sitting at $p_{{}_{\!G}}$, this explicit time 
parameter is the value of the scalar field $X^0$, that is, the intrinsic time.

\vspace{4mm}

One may regard expression (\ref{evolv}) for the invariants as an
expression for ``evolving constants of motion" since  it can be read
as a combination of constants of motion with coefficients - the
monomials $t^n$ - changing in time. As a matter of fact, expression
(\ref{evolv}) is an explicit realization of the approach
\cite{Rov1,Rovelli:1989jn,Rovelli:1990jm}, which is often referred
to as the program of ``evolving constants of motion". We believe
that the clarifications made above, although restricted to classical
considerations, put in a new and satisfying perspective the concept
of  ``evolving constants of motion". This terminology was adopted in
order to circumvent a problem which from our perspective never
existed in the first place - at least in the classical setting. The
problem - the notion that ``nothing happens'' in generally covariant
theories'' - arises from a failure to adequately distinguish between
time evolution and gauge symmetry. We devote Appendix A to this
analysis. We think that our contribution makes superfluous this
terminology, although the idea behind it remains fully vindicated.
The resolution by Rovelli of the supposedly apparent paradox amounts
to \cite{Anderson:1995tu} ``the recognition that observables are
members of families of constants of motion parametrized by a label
related to time".  In other words, as the time coordinate 
for a particular observer evolves and takes different values, 
so do the observables.  The observables consist of a sum - perhaps even of an infinite number of terms - of constants of the motion multiplying increasing powers of the time coordinate. These constant coefficients are expressed as invariant functionals of the phase space variables. The coordinate time itself is of course not a canonical variable, and is therefore invariant under the action of the canonical gauge group. There has been a subsequent effort
by a list of authors, particularly \cite{rovelli02,thiemann06}, to
obtain a consistent quantum picture for the observables. In generic
spacetimes it will be necessary to restrict to locally defined
intrinsic time. One must also address the question of equivalence of
quantum theories based on different choices.

\vspace{4mm}

\subsubsection{From observables to generators of rigid Noether symmetries}
\label{getnoether}

 Consider the observables associated with fields other than lapse and shift. The constants of motion obtained from these  observables are Noether generators of symmetries for the reduced phase space where lapse, shift and their canonical momenta have been eliminated. If we want to construct Noether generators for the entire phase space, one can proceed as follows. Let ${\cal C}$ be one of such constants of motion. It is a functional of the
fields $\Phi(\vec x)$ - other than lapse and shift - and it exhibits explicit dependence on the
spatial coordinates $\vec x$ as well, through $\chi^i= x^i -X^i$. Its weakly vanishing Poisson bracket
with the Hamiltonian constraints can be expressed as
$$\{{\cal C},\,{\cal H}_\mu \}= U_{\ \mu}^\nu {\cal H}_{\ \nu}\,,
$$
for some functional matrix $U_{\ \mu}^\nu$. Then the following construction
$$
{\cal Q} := {\cal C} +  U_{\ \mu}^\nu N^\mu P_\nu\,,
$$ satisfies ($H_c$ stands for Dirac's canonical Hamiltonian, $H_c= N^\mu{\cal H}_\mu$)
$$
\frac{\partial {\cal Q}}{\partial t}+ \{{\cal Q},\,H_c \}= 0 + \{{\cal Q},\,N^\mu{\cal H}_\mu \}
= {\cal O}(P)\,,
$$
and
$$
 \{{\cal Q},\,P_\mu \}= {\cal O}(P)\,,
$$
(where ${\cal O}(P)$ means terms linear in the momentum constraints) which are the two conditions
spelled out in \cite{Batlle:1987ek} for ${\cal Q}$ to qualify as a Noether conserved quantity
associated with a symmetry which is projectable from tangent space to phase space. In fact this
symmetry is generated by ${\cal Q}$ through the Poisson bracket\footnote{Note in particular that
${\cal Q}$ generates variations for the lapse and shift according to
$\delta N^\mu = \{N^\mu ,\,{\cal Q} \} = U_{\ \nu}^\mu N^\nu\,.$}.

Notice that these Noether generators ${\cal Q}$ are gauge invariant
quantities because their Possion bracket with the full set of
primary and secondary constraints weakly vanishes. In practice, to
smooth out the dependence of ${\cal Q}$ on the spatial coordinates,
one can use arbitrary smearing functions $\alpha(\vec x)$ and define
generators $${\cal Q}_{\!{}_\alpha} := \int d^3x \,\alpha(\vec
x)\,Q(\vec x)\,.$$ Let us stress that ${\cal Q}_{\!{}_\alpha}$ are
generators of {\sl rigid} Noether symmetries. To be gauge symmetries
one should allow the functions $\alpha$ to have arbitrary dependence
on the time coordinate, but if we allow that, then  ${\cal
Q}_{\!{}_\alpha}$ ceases to be a constant of motion and a Noether
generator.

\vspace{4mm}

 As regards the constants of motion extracted from the observables associated with the lapse and shift, one should consider ${\cal I}_{{}_{\!N^\sigma}}  = exp\left(  \{ - , \,G_{\overline\xi }\} \right)N^\sigma$, with $G_{\overline\xi }$ as in (\ref{gengen}). This gives an expansion
$$ {\cal I}_{{}_{\!N^\sigma}} = N^\sigma +
\Big(\{N^\sigma,\,{\overline P}_\nu\dot{\overline\xi^\nu}\} + \ldots\Big)_{{\overline \xi}^\sigma \to
\chi^\sigma,\  {\dot{\overline\xi^\sigma}}\to{\dot{\chi^\sigma}}} = {\cal B}_0^\sigma+ \ldots\,,
$$
where we have used $\dot{\chi^\sigma}= \delta^\sigma_0-N^\nu {\cal A}^\sigma_{\ \nu}$. In fact there is a
quick shortcut to sum this expansion, because we know that in $p_{{}_{\!G}}$ the - secondary -  gauge
fixing constraints are satisfied, i.e., $\delta^\sigma_0-\hat N^\nu\hat {\cal A}^\sigma_{\ \nu}=0$, and
therefore $\hat N^\sigma= {\cal I}_{{}_{\!N^\sigma}} = {\cal I}_{{}_{{\cal B}_0^\sigma}}$. Since ${\cal
B}_0^\sigma$ depends on fields other than lapse and shift, we can conclude that using the mechanism
explained above in this subsection, the constants of motion extracted from {\sl all} our invariants can be
made Noether symmetry generators.

\vspace{4mm}

Finally, let us discuss one more basic aspect of the rigid symmetries generated by
${\cal Q}_{\!{}_\alpha}$: they move an on shell field configuration out of its gauge orbit. In fact,
since they commute with the gauge generators, they will map an entire gauge orbit into another.
To prove that indeed  ${\cal Q}_{\!{}_\alpha}$ moves a field configuration out of the gauge orbit
we only need to verify that the variations generated by ${\cal Q}_{\!{}_\alpha}$ through the Poisson
bracket do not leave invariant the set of gauge invariant functionals ${\cal I}_{\Phi}$. This is proven
in the next section \ref{secdirac}, where we compute the Poisson bracket algebra of the invariants
and obtain a non-trivial result.

\subsubsection{Interpreting the constants of motion}
\label{interct}

What has been done for the  time coordinate may be done for
any other  coordinate. Recall that the invariant associated
with the field $\phi$ is
$${\cal I}_\phi \approx \sum_{n=0}^{\infty} \frac{1}{n!}\chi^n \{\phi,\,{\cal H}\}_{(n)}\,,
$$
with $\chi = x - X$\,. When the explicit time coordinate is fixed at an arbitrary value,
$I_\phi$ becomes a constant of motion. 

Applying arguments similar to those leading up to (\ref{dynob}) we
find that \beq\frac{d}{d\,x^a} {\cal I}_{\Phi}\approx{\cal I}_{\{
\Phi , \,\overline{\cal H}_a\}}\,. \label{xaderivob}\eeq Recall that
the invariant  ${\cal I}_{\Phi}$ is the  field $\Phi$ 
evaluated at the gauge-fixed point $p_G$,  ${\cal I}_{\Phi}= \hat\Phi$. 
To find $\overline{\cal H}_a$ at $p_G$ we need to make use of
the fact that the $X^\mu$ are  scalars under spatial coordinate
transformations, i.e., under $\bar x^\mu = x^\mu - \xi^a  \delta^\mu_a$,
$$
 \delta X^\mu =  \left\{ X^\mu,  \int d^3\!x\,\xi^a {\cal H}_a \right\} = X^\mu_{,a} \xi^a,
$$
and therefore ${\cal A}^\mu_a = X^\mu_{,a}$. As a consequence $\left.  {\cal A}^\mu_a\right|_{p_G} = \delta^\mu_a$.
Taking into account that according to (\ref{nmu}) $\left.  {\cal B}^\mu_0\right|_{p_G} = N^\mu$ we find that
$$
\left.  {\cal B}^\mu_a\right|_{p_G} = \delta^\mu_a.
$$
It follows finally that
\beq
{\cal I}_{\{ \Phi , \,\overline{\cal H}_a\}} = \left. \frac{\partial \Phi}{\partial x^a}\right|_{p_G}. \label{derphi}
\eeq

We know that due to gauge invariance, there is no implicit
dependence on time (when the invariant is evaluated on an on shell
configuration, a solution of the equations of motion). For the same
reason, there is no implicit dependence on the spatial coordinates
either since spatially constant translations are gauge
transformations. Let us write ${\cal I}_\Phi$ in powers of
all explicit coordinates, 
\beq {\cal I}_\phi \approx \sum_{n_\mu=0}^{\infty}
\frac{1}{n_0!\,n_1!\,n_2!\,n_3!}\,
(x^0)^{n_0}(x^1)^{n_1}(x^2)^{n_2}(x^3)^{n_3}\, {\cal
C}_{n_0,n_1,n_2,n_3}\ , 
\eeq 
with 
\bea {\cal C}_{n_0,n_1,n_2,n_3}:=
{\cal I}_{\{\{\{\{ \phi,\,,\overline{\cal H}_0
\}_{(n_0)}\,,\overline{\cal H}_1\}_{(n_1)}\,,\overline{\cal
H}_2\}_{(n_2)}\,,\overline{\cal
H}_3\}_{(n_3)}}{}_{\!\Big\vert_{x^\mu=\,0}}\,. \label{4d} 
\eea (The
order of the Hamiltonians is irrelevant owing to the strongly
vanishing Poisson bracket property.) Notice that setting $x^\mu=\,0$
in (\ref{4d}) refers only to the explicit coordinate dependencies.
Indeed, the coefficients ${\cal C}_{n_0,n_1,n_2,n_3}$ may be
evaluated in any arbitrary system of coordinates; they are
invariants. Consider, for example, that we are in a point $p$ in
$\cal S$. To evaluate the value of the functional ${\cal
C}_{n_0\,n_1\,n_2\,n_3}$ we just substitute in the field
configurations $\phi_p^A(x)$ and an indefinite number of their
spatial derivatives - which appear due to the nesting of Poisson
brackets - in (\ref{4d}). The result does not depend on the time
coordinate $x^0$ nor on the spatial coordinates $\vec x$ where all
the fields and their spatial derivatives are computed. If instead of
inserting $\phi_p^A(x)\,,\partial_i
\phi_p^A(x)\,,\partial_i\partial_j \phi_p^A(x)... \forall A$, we
were to substitute in the on shell field configuration at another
point $q$ in $\cal S$ and with other values $y$ of the coordinates,
$\phi_q^A(y),\partial_i \phi_q^A(y)$, the numerical result would be
the same.

Generalizing (\ref{derphi}), we deduce that these constants of
motion ${\cal C}_{n_0,n_1,n_2,n_3}$ are the values of the
corresponding $n_\mu$ partial derivatives of $\hat \phi$ at the zero
value of the intrinsic coordinates. (In fact one could have expanded
around any other values.) Thus the formalism manages to pick data at
any time, $t$, at some point $p$ in $\cal S$, and to convert these
data into the coefficients of the Taylor expansion for the fields at
$p_{{}_{\!G}}$, which are obviously invariants. By the same token,
the information in $I_\phi$ itself is that of the field
configuration at $p_G$ at the given time that appears in the
invariant as the explicit time. In the particular case of a scalar
field and for an observer sitting at $p$, $I_{\psi}(t, \vec x)$ it is
the value of the field at the very moment and place where the
gauge-fixed scalars  $X^\mu$ take the values $t, \vec x$.

\vspace{5mm}

{\bf A simple example}

\vspace{3mm}

Now we demonstrate with the simplest of the examples, that of the massive free particle in Minkowski 
spacetime, our findings above\footnote{ The example of a dimensionally reduced spatially
homogeneous isotropic cosmological model is worked out in \cite{Pons:2009cd}.}. Consider the Lagrangian
$$ L = \frac{1}{2\,N}\eta_{\mu\nu}\dot q^\mu\dot q^\nu - \frac{1}{2} m^2 N,
$$
where $N$ is an auxiliary variable - in fact it is the lapse -, and $\eta^{\mu\nu}= (-,+,+,+)$. The Dirac Hamiltonian is
$$ H_{\!D} = \frac{1}{2} N (\eta^{\mu\nu} p_\mu p_\nu +  m^2) + \lambda \pi,
$$ 
where $\pi$, the momentum canonically conjugate to $N$, is the primary constraint and $\lambda$ an arbitrary function of time. There is a secondary constraint, namely ${\cal H} = \frac{1}{2} (\eta^{\mu\nu} p_\mu p_\nu + m^2)$.
The gauge generator has the form $G = \xi {\cal H} + \dot \xi \pi$. We choose as a gauge fixing constraint $\chi = t-q^0$. Next, following the instructions in section \ref{descrip}, we define 
$A:= \{q^0,\,{\cal H}\} = p^0$, and 
$$\overline{\cal H} = \frac{1}{A}{\cal H} =\frac{1}{2\,p^0}(\eta^{\mu\nu} p_\mu p_\nu + m^2)\,.
$$ 
Now we are ready to compute the invariants. Note that we do not write the implicit time dependence in the variables, which is the same as if we were working just in phase space instead of working in the space of trajectories - i.e., field configurations. The series expansions are trivial and we get
\bea {\cal I}_{q^i} &=& q^i + \chi \{q^i,\,\overline{\cal H}\} = q^i + (t-q^0) \frac{p^i}{p^0} =
  (q^i-\frac{p^i}{p^0}q^0) + \frac{p^i}{p^0} t,\nonumber \\ {\cal I}_{p_\mu} &=& p_\mu\,. \nonumber 
\eea
Thus we identify from the expansion in the $t$ parameter for ${\cal I}_{q^i}$ the constants of motion $q^i-\frac{p^i}{p^0}q^0$ and $\frac{p^i}{p^0}$, and from ${\cal I}_{p_\mu}$ the constants of motion $p_\mu$.
We have seven independent constants of motion that can be written $p_\mu$ and $c^i := p^0 q^i - q^0 p^i$. These are the  Poincar\`e translation and boost generators. Note that the combinations $\frac{1}{p^0}(p^i c^j -p^j c^i)$ of these constants of motion
are $p^i q^j - p^j q^i$, that is, the generators of rotations. The full Poincar\'e algebra of generators of rigid symmetries of the free particle is obtained.

Finally, using the methods introduced in section \ref{anyfield}, one can compute the invariant associated with the lapse $N$. The result is  ${\cal I}_{N}= \frac{1}{p^0}$, that is, one of the constants of motion obtained  above.

\subsection{Observables and Dirac brackets}
\label{secdirac}

\subsubsection{Preliminary remarks}

It is a remarkable fact that for some purposes the explicit
construction of invariants for which the general theory has been
given above can be avoided. In this subsection we will show that the
Poisson bracket $\{{\cal I}_{\Phi^A},\,{\cal I}_{\Phi^B}\}$ of the
invariants associated with the fields $\Phi^A,
\Phi^B$, is the invariant associated with the Dirac
bracket of the fields themselves.

As a preliminary observation, one might wonder how we can compute
Poisson brackets of the functional invariants ${\cal I}_{\Phi^A}$,
given that their arguments are only defined for fields satisfying
the equations of motion; Poisson brackets involve arbitrary
variations, including ``off shell'', i.e., violating the equations
of motion.  The resolution  is the following. Since the Poisson
brackets are an equal-time computation, let us simply examine the
functionals at a given time $t$. An arbitrary extension {\sl off
shell} of a functional ${\cal I}_{\Phi^A}$ will produce ${\cal
I}_{\Phi^A}\rightarrow{\cal I}_{\Phi^A} +{\cal O}(P_\mu,\, {\cal
H}_\nu)$. But notice that this {\sl off shell} extension does not
change the Poisson brackets as long as we evaluate the result on
shell since $\{{\cal O}(P_\mu,\, {\cal H}_\nu),\,{\cal F}_\Phi\}
\approx 0$  and also of course $\{{\cal O}(P_\mu,\, {\cal
H}_\nu),\,{\cal O}(P_\rho,\, {\cal H}_\sigma)\} \approx 0$. So
indeed the functionals need only be defined on shell for their
Poisson bracket to be well defined on shell.

We will show that
$$\{{\cal I}_{\Phi^A},\,{\cal I}_{\Phi^B}\} \approx {\cal I}_{\{\Phi^A,\,\Phi^B\}^*}\,,
$$
where the Dirac bracket is
$$
\{\Phi^A,\,\Phi^B\}^* := \{\Phi^A,\,\Phi^B\} - \{\Phi^A,\, C^i\} M^{-1}_{i j} \{C^j,\,\Phi^B\}.
$$
In this expression we define the eight member set $C^i := {\cal H}_\mu,\, \chi^\nu $,
and $M^{i j} := \{ C^i, C^j \}$.

Notice that if the map $\Phi^A\rightarrow{\cal I}_{\Phi^A}$ were a
canonical transformation, the result would have been simply $\{{\cal
I}_{\Phi^A},\,{\cal I}_{\Phi^B}\} = {\cal
I}_{\{\Phi^A,\,\Phi^B\}}\,, $ because the Poisson bracket structure
is preserved by a canonical transformation. The crucial fact that
complicates this computation is that the descriptors, which are
determined by the gauge fixing conditions, are substituted by
functionals of the field configurations at $p$ {\sl after} the
action of the finite element of the gauge group $exp(\{-,\,G\})$ is
taken, as it is clear in (\ref{expansion}). We further discuss in
\ref{limits} this issue of non-canonicity of the map
$\Phi^A\rightarrow{\cal I}_{\Phi^A}$.

\vspace{4mm}

The following proof is restricted, just for simplicity, to canonical fields other than the
lapse and shift and their conjugates.  Thus for the generator
(\ref{thegen})  we only need the reduced expression $G= {\bar
\xi}^{\mu} {\bar {\cal H}}_{\mu}$, with ${\bar {\cal H}}_{\mu}$ defined in (\ref{hbarra}). This 
restriction  is easily elimininated by taking into account the results in section (\ref{anyfield}) 
and using the generator (\ref{gengen}). The proof will be undertaken in two steps.
In the first step we show that this relation holds at
$p_{{}_{\!G}}$, and in the second step the proof is extended to
an arbitrary point in the gauge orbit.

\subsubsection{Proof - Step 1: Neighborhood of $p_{{}_{\!G}}$}

Let us consider a neighborhood of $p_{{}_{\!G}}$ in the gauge orbit, and take an
arbitrary point $p$ in the same orbit, such that  the set of descriptors
 used to bring configurations in $p$ to configurations
in $p_{{}_{\!G}}$ are infinitesimal.
Let us write, recalling (\ref{expansion}) and keeping terms to first order in the
infinitesimal  descriptors,
$$
{\cal I}_{\Phi^A}   = \Phi^A + \chi^\mu\{\Phi^A,\,{\bar {\cal H}}_\mu\} + {\cal O}(\chi)^2.
$$
Then, computing at $p$,
\bea
\{{\cal I}_{\Phi^A},\,{\cal I}_{\Phi^A}\} &=& \{\Phi^A
+ \chi^\mu\{\Phi^A,\,{\bar {\cal H}}_\mu\},\,\Phi^B
+ \chi^\nu\{\Phi^B,\,{\bar {\cal H}}_\nu\}\}\nonumber \\
&=&
\{\Phi^A,\,\Phi^B\} - \{\Phi^A,\,{\bar {\cal H}}_\mu\}\{\chi^\mu,\,\chi^\nu\}
\{{\bar {\cal H}}_\nu,\,\Phi^B\}\nonumber \\ &-&
 \{\Phi^A,\,\chi^\mu\}\{{\bar {\cal H}}_\mu,\,\Phi^B\}
+ \{\Phi^A,\,{\bar {\cal H}}_\mu\}\{\chi^\mu,\,\Phi^B\} \nonumber\\
&+& {\cal O}(\chi)
\nonumber \\&=& \{\Phi^A,\,\Phi^B\}^* + {\cal O}(\chi)\,.
\label{DB1}
\eea
In the last equality we have used the fact that
$$M = \left(\begin{array}{cc} \{{\bar {\cal H}},\,{\bar {\cal H}}\} &
\{{\bar {\cal H}},\,\chi\} \\
\{\chi,\,{\bar {\cal H}}\} & \{\chi,\,\chi\} \end{array}\right) \approx
\left(\begin{array}{cc} 0 &
+\delta \\
-\delta & \{\chi,\,\chi\} \end{array}\right)\,,
$$
(our fields satisfy the equations of motion, so
$\{{\bar {\cal H}},\,{\bar {\cal H}}\} \approx 0$\footnote{ In Appendix B we derive the stronger
result $\{{\bar {\cal H}},\,{\bar {\cal H}}\} = {\cal O}({\cal H}^2)$.})
has as its inverse
$$M^{-1} =
\left(\begin{array}{cc} \{\chi,\,\chi\} &
-\delta \\
+\delta & 0 \end{array}\right)\,,
$$
thus producing the Dirac brackets above. This computation has been
made at $p$, in the close neighborhood of $p_{{}_{\!G}}$. Now we can
take   the limit $p\rightarrow p_{{}_{\!G}}$ on both sides, thus
obtaining 
\beq 
\{{\cal I}_{\Phi^A},\,{\cal
I}_{\Phi^B}\}{}_{\big|_{p_{{}_{\!G}}}} =
\{\Phi^A,\,\Phi^B\}^*{}_{\big|_{p_{{}_{\!G}}}}\,={\cal
I}_{\{\Phi^A,\,\Phi^B\}^*}
{}_{\big|_{p_{{}_{\!G}}}}\,,
\label{DB2} 
\eeq 
 where in the last step we have used the fact that
the functionals ${\cal I}_{\Phi^A}$ become the identity functionals
when their arguments are taken at $p_{{}_{\!G}}$. This
concludes the first step of our proof.

\vspace{4mm}

An alternative proof, using the connection between the observables
and specific canonical transformations of the fields is given in
section \ref{revisitdb}.

\subsubsection{Proof - Step 2: Gauge orbit}
Let us now extend this result to the entire gauge orbit. We can make
an arbitrary gauge transformation sending  the
equality(\ref{DB2}) holding at $p_{{}_{\!G}}$ to a
corresponding equality at any other point $p$. Let us call
$U(p,p_{{}_{\!G}})$ this gauge transformation from $p_{{}_{\!G}}$ to
$p$. Its specific descriptors can be determined in a manner similar
to the procedure for building the invariants. $U(p,p_{{}_{\!G}})$ is
a canonical transformation and, as such, preserves the Poisson
bracket structure. This means that $U(p,p_{{}_{\!G}})$ ``enters" on
both sides of the Poisson bracket. On the other hand, the action of
$U(p,p_{{}_{\!G}})$ on the functionals is
$${\cal I}_{\Phi^A}[X^\mu_{p_{{}_{\!G}}}, \Phi^A_{p_{{}_{\!G}}};\,x]
\rightarrow {\cal I}_{\Phi^A}[X^\mu_p, \Phi^A_p;\,x]\,.
$$
(In fact the action of $U(p,p_{{}_{\!G}})$ on these functionals is trivial because they
are invariant under the gauge transformations and can be written in terms of
the fields at any point in the gauge orbit). These considerations show that
the left hand side of (\ref{DB2}) undergoes, under the action of $U(p,p_{{}_{\!G}})$, the
transformation
$\{{\cal I}_{\Phi^A},\,{\cal I}_{\Phi^B}\}{}_{\big|_{p_{G}}}\rightarrow
\{{\cal I}_{\Phi^A},\,{\cal I}_{\Phi^B}\}{}_{\big|_{p}}$.

Let us now address the transformation of the right hand side of (\ref{DB2}) under $U(p,p_{{}_{\!G}})$.
If the canonical gauge transformation
$U(p,p_{{}_{\!G}})$ could ``enter" within the Dirac brackets, then the result for the
transformation of the right hand side would be simply
$\{\Phi^A,\,\Phi^B\}^*{}_{\big|_{p_{G}}}\rightarrow \{\Phi^A,\,\Phi^B\}^*{}_{\big|_{p}}$. But the
Dirac bracket
structure is not preserved by canonical transformations generated by
the Poisson bracket since $\{U\,\Phi^A,\,U\,\Phi^B\}^* \neq U\,\{\Phi^A,\,\Phi^B\}^*$.
Indeed, we can now take advantage of writing $\{\Phi^A,\,\Phi^B\}^*{}_{\big|_{p_{G}}}$
as ${\cal I}_{\{\Phi^A,\,\Phi^B\}^*}{}_{\big|_{p_{G}}}$ because then
$\{\Phi^A,\,\Phi^B\}^*$ becomes just the label that identifies the functional
we are considering, in the sense that ${\cal I}_{\{\Phi^A,\,\Phi^B\}^*}$ is the functional that
sends the specific combination of field configurations given by $\{\Phi^A,\,\Phi^B\}^*$ to its
value at $p_{{}_{\!G}}$. Thus it is obvious that the action of $U(p,p_{{}_{\!G}})$ on this functional
just maps $${\cal I}_{\{\Phi^A,\,\Phi^B\}^*}{}_{\big|_{p_{G}}}\rightarrow
{\cal I}_{\{\Phi^A,\,\Phi^B\}^*}{}_{\big|_{p}}
$$

The equality between the transformed objects in the left hand side and the right hand side of (\ref{DB2})
tells us that we have obtained
\bea
\{{\cal I}_{\Phi^A},\,{\cal I}_{\Phi^B}\} \approx {\cal I}_{\{\Phi^A,\,\Phi^B\}^*}\,,
\label{DB3}
\eea
for any arbitrary point $p$ in the gauge orbit.

\subsubsection{Additional remarks}

The results in section \ref{anyfield} permit the extension of
 (\ref{DB3}) to the lapse and shift fields. Indeed the
situation is the same as in section \ref{secdirac} when working with fields other than lapse and
shift, but instead of having $4$ first class constraints and $4$ gauge fixing constraints, there
are now $8$ constraints of each type. Proceeding exactly through the same steps as in
section  \ref{secdirac}, the result (\ref {DB3}) can be extended to all the fields, lapse and
shift included.

The result (\ref{DB3}) has been previously obtained by Thiemann
\cite{thiemann06} in a remarkable proof based on formal series
expansion or - in his own words - by ``brutally working out the
Poisson brackets." We have provided a natural geometric interpretation of this series expansion.

 So far our
considerations hold for ${\cal S}$, the space of on-shell field
configurations.  But once the results have been obtained, and
recalling that all the canonical gauge transformations are active
transformations at fixed spacetime coordinates, we can examine all
our actions along the gauge orbit at a fixed value of the time
coordinate, $t_0$. At this fixed time, which can be considered the
time for the setting of the initial conditions, the field
configurations only need to satisfy the constraints $P_\mu \approx
0, \  {\cal H}_{\mu}\approx 0$. Thus our results are valid in a
phase space formulation on the entire original first class
constraint surface (but not including the gauge fixing constraints,
which are only satisfied at the particular point $p_{{}_{\!G}}$ in
the gauge orbit).

\vspace{4mm}

We notice also that the results obtained above are different
from the results in \cite{henneauxt94}
showing that the Dirac bracket of the invariant functionals
coincides on shell with its Poisson bracket; see especially
Exercise 1.18 and section 13.2.2 in this book. This is obviously
true by the very nature of the invariant functionals; they are
required  to satisfy $\{{\cal I}_{\Phi^A},\,G\}\approx
0$\footnote{At a fixed time $t_0$ this reads $\{{\cal
I}_{\Phi^A},\,P_\mu\}\approx 0,\ \{{\cal I}_{\Phi^A},\,{\cal
H}_\mu\}\approx 0$.}, which is the ingredient needed to show, in
the light of the Dirac bracket (\ref{DB1}), or its generalization to
all $8+8$ constraints, that indeed
$$\{{\cal I}_{\Phi^A},\,{\cal I}_{\Phi^B}\}^* \approx \{{\cal I}_{\Phi^A},\,{\cal I}_{\Phi^B}\}\,.
$$
Taking (\ref{DB3}) into account, and including the result above, we
can write in phase space, \bea \{{\cal I}_{\Phi^A},\,{\cal
I}_{\Phi^B}\}^*  \approx {\cal I}_{\{\Phi^A,\,\Phi^B\}^*}\,
\label{DB4} \eea where we have expressed the fact, using  the weak
equalities ``$\approx$''  that these relations are satisfied on the
constraint hypersurface surface in phase space - again, not
including the gauge fixing constraints.

\subsection{The invariants constructed as limits of canonical maps}
\label{limits}

\subsubsection{On the non-canonicity of the map $\Phi\rightarrow{\cal I}_{\Phi}$}

To further study some aspects of the
observables , we will elaborate on the non-canonicity of the map
$\Phi\rightarrow{\cal I}_{\Phi}$ and its proximity to canonical
maps. In order not to overload the subsequent considerations we
exclude the lapse and the shift fields, but remark that by the
techniques described in (IIB3) the results of this
 subsection can be extended to lapse and
shift.

Let us recall the expression (\ref{expansion}) for the invariant
functional ${\cal I}_{\Phi}$, having chosen the usual basis for the
Hamiltonian constraints such that, $\{
{\chi}^{\nu},\, \overline{\cal H}_{\mu}\}=-\delta^{\nu}_{\mu} + {\cal
O}(\overline{\cal H})$ and $\{ \overline{\cal H}_{\nu},\, \overline{\cal H}_{\mu}\}={\cal
O}(\overline{\cal H}^2)$. Let us write again the expression for our
observables, after equation (\ref{expansion}),
\beq  {\cal I}_{\Phi} := exp\left(  \{ -
,\,\overline{\xi}^{\nu} \overline{\cal H}_{\nu}\} \right)\Phi{}_{\big|_{\xi=\chi}}
\approx \sum_{n=0}^{\infty}\frac{1}{n!}\chi^n\{ \Phi , \,
 \overline{\cal H}\}_{(n)}\,.
\label{expansion2}
\eeq

\vspace{4mm}

The map $\Phi\rightarrow{\cal I}_{\Phi}$ sends all the points $p$ in
the gauge orbit to a single point $p_{{}_{\!G}}$. Hence it can not
be a canonical transformation because one such transformation should
be invertible. An alternative, indirect but sufficient proof of this
non-canonicity is that the Poisson bracket of the invariants
associated with two given fields is not the invariant associated
with the Poisson bracket of these fields, but with the Dirac
bracket. The reason for this non-canonicity may be traced to the
fact that the descriptors ${\xi}^{\nu}$ are replaced by the gauge
fixing constraints $\chi^\mu$ {\sl after} the action of the finite
gauge transformation that sends $p$ to $p_{{}_{\!G}}$.

We will explore how close this map $\Phi\rightarrow{\cal I}_{\Phi}$
can be to a canonical transformation. We will show that it is in
fact the limit of a family of canonical transformations. To
construct this family, an obvious candidate is the object that
results from making the replacement of the descriptors {\sl before}
the action of the finite gauge transformation. A one-parameter
family of canonical transformations is found by allowing a global
rescaling for the descriptors.

\vspace{4mm}

So consider the functional, for $G:= {\chi}^{\nu} \overline{\cal H}_{\nu}$,
\bea{\cal K}_{\Phi} &:=& exp\left(  \{ - ,\lambda\,G\} \right)\Phi
= exp\left(  \{ - ,\lambda\,{\chi}^{\nu} \overline{\cal H}_{\nu}\} \right)\Phi \nonumber \\
&=& exp\left(  \{ - ,\lambda\,{\chi} \overline{\cal H}\} \right)\Phi\,,
\label{defK}
\eea
with $\lambda$ a real parameter. Thus the map $\Phi\rightarrow{\cal K}_{\Phi}$ is canonical.
We will show that ${\cal I}_{\Phi}$ can be reobtained as the $\lambda\to\infty$ limit of ${\cal K}_{\Phi}$.

\vspace{4mm}

We start with the expansion for ${\cal K}_{\Phi}$,
\beq {\cal K}_{\Phi} =exp\left(  \{ - ,\lambda\,G\} \right)\Phi
=
\sum_{n=0}^{\infty} \frac{\lambda^n}{n!} \{ \Phi , \,
 G\}_{(n)}\,.
\label{expG} \eeq To continue, let us define $B_n:=\chi^n \{ \Phi ,
\,\overline{\cal H}\}_{(n)},\, n>0$, with $B_0:=\Phi $. Our aim is
to rewrite the expansion (\ref{expG}) in terms of these objects
$B_n$. Notice that, due to the fact that $\{ \chi ,\overline{\cal
H}\} \approx -\delta$, we get the simple relation
\beq\{ B_n , \,
 G\}\approx -n B_n + B_{n+1}\,.
\label{simple}
\eeq
Our result will take the form
\bea
{\cal K}_{\Phi} \approx \sum_{n=0}^{\infty} c_n \,B_n\,,
\label{ksumg}
\eea
and the task is to compute the coefficients $c_n$.

\vspace{4mm}

One can see immediately that $c_0=1$. To compute $c_1$ we need to
add all the appearances of $B_1$ in the different terms in
(\ref{expG}). We find, keeping only the $B_1$ terms,
$${\cal K}_{\Phi}
\approx
\Phi +  \lambda B_1  +\frac{\lambda^2}{2!}(-B_1+ \ldots)+\frac{\lambda^3}{3!} (B_1+ \ldots) +\ldots
$$
and thus
$$
c_1= \lambda-\frac{\lambda^2}{2!}+\frac{\lambda^3}{3!} +\ldots = 1-e^{-\lambda}\,.
$$

\vspace{4mm}

It turns out that
\beq
c_n=\frac{1}{n!}\,(1-e^{-\lambda})^n\,.
\label{cn}
\eeq
 This result will be obtained below employing a different technique.

\vspace{4mm}

With the coefficients (\ref{cn}) we obtain, for (\ref{ksumg}),
\beq {\cal K}_{\Phi}
\approx \sum_{n=0}^{\infty}\frac{1}{n!}(1-e^{-\lambda})^n \chi^n \{ \Phi , \,
 \overline{\cal H}\}_{(n)}\,,
\label{theK}
\eeq

It is illuminating to notice the substantial difference between the two series expansions, (\ref{expG})
and (\ref{theK}), for the same functional ${\cal K}_{\Phi}$. Both are power series expansions but
whereas the first is in terms of the parameter $\lambda$, the second is in terms of
$(1-e^{-\lambda})$ and is only valid on shell. One reasonably expects convergence at least in the
case where the point $p$ in the space of field configurations $\cal S$ is in the neighborhood
of $p_{{}_{\!G}}$.

\vspace{4mm}

Notice that, as expected, there is no finite $\lambda$ that can make
${\cal K}_{\Phi}={\cal I}_{\Phi}$. Curiously
enough, though, and as long as it is legitimate to enter the limit $\lambda\rightarrow\infty$ within
the series expansion (\ref{theK}), one finds, recalling (\ref{expansion2}),
$$\lim_{\lambda\rightarrow\infty}{\cal K}_{\Phi} \approx {\cal I}_{\Phi}\,,
$$
thus the invariants can be interpreted as limits of one-parameter
families of canonical transformations. But such a limit is no longer
a canonical transformation\footnote{Compare with the homothetic map
${\mathcal R}^2 \rightarrow {\mathcal R}^2$ defined by ${\vec
v}\rightarrow\frac{1}{\lambda}{\vec v}$, which is invertible for any
real value of ${\lambda}$. It becomes singular for
$\lambda\rightarrow\infty$: \ ${\vec v}\rightarrow{\vec 0}$.}. Our
previous analysis makes the reason more transparent because whereas
the limit $\lambda\rightarrow\infty$ can be easily taken for the
expansion (\ref{theK}), for which it simply says
$(1-e^{-\lambda})\rightarrow 1$, it makes no sense at all for
(\ref{expG}). And here is the point: it would have been  just by
obtaining a finite result for the computation of the limit in the
exponent $  \{ - ,\lambda\,G\}  $ in (\ref{expG}), which is clearly
divergent, that we would have been assured that the end result was a
canonical transformation.

\vspace{4mm}

A complementary result is obtained by considering the computation of
$\{{\cal K}_{\Phi},\,\overline{\cal H}\}$. It is crucial in this
regard that the Poisson bracket of the constraints $\overline{\cal
H}$ among themselves is quadratic in the constraints (see Appendix
B). Owing to this fact, $  \{ G ,\,\overline{\cal H}\} = -
\overline{\cal H} + {\cal O}(\overline{\cal H}^2)$, and the
quadratic terms can be dropped in the internal Poisson brackets as
long the final result is expressed on shell. Taking into account
(\ref{expG}) and that
$$
\{\{\Phi,\,G\}_{(n)},\,{\cal H}\} \approx
\sum_{k=0}^{n}(-1)^{(n-k)}\frac{n!}{k!\,(n-k)!}\{\{\Phi,\,{\cal H}\},\,G\}_{(k)}\,,
$$
one easily obtains\footnote{This result can also be obtained by
differentiating with respect to the parameter $\lambda$ the on shell
equivalent expressions (\ref{expG}) and (\ref{theK}), but here we
choose another method, based exclusively on (\ref{expG}), in order
to provide a proof of (\ref{theK}).} \beq\{{\cal
K}_{\Phi},\,\overline{\cal H}\} \approx e^{-\lambda} {\cal
K}_{\{\Phi,\,\overline{\cal H}\}}\,,\label{kh} \eeq which, in the
limit $\lambda\rightarrow\infty$, tells us again that ${\cal
I}_{\Phi}$ is an invariant, that is, $\{{\cal
I}_{\Phi},\,\overline{\cal H}\}\approx 0$.

Notice that we can use (\ref{kh}) to obtain the coefficients
(\ref{cn}). Consider the generic expansion (\ref{ksumg}) for ${\cal
K}_{\Phi}$ and require that it complies with (\ref{kh}). One easily
finds a recurrent equation for the coefficients $c_n$,
$$ c_n - (n+1)c_{n+1} = e^{-\lambda} c_n\,,$$
which, with the obvious input $c_0=1$, yields the result (\ref{cn}).

\subsubsection{Revisiting the Dirac bracket}
\label{revisitdb}

The  considerations  in this
subsection  provide an alternative computation of the
Poisson bracket of the invariants  as compared to (IIIB).
From the definition (\ref{defK}) it is clear that, since
${\Phi}\rightarrow {\cal K}_{\Phi}$ is a canonical transformation,
$$\{{\cal K}_{\Phi^A},\,{\cal K}_{\Phi^B}\} = {\cal K}_{\{\Phi^A,\,\Phi^B\}}\,.
$$

The fact that (\ref{kh}) implies in general that $\{{\cal
K}_{\Phi},\,\overline{\cal H}\}\neq 0$, means that to compute the
Poisson bracket $\{{\cal K}_{\Phi^A},\,{\cal K}_{\Phi^B}\}$ off
shell information of ${\cal K}_{\Phi}$ must be used. The minimal off
shell information we need is ${\cal O}(\overline{\cal H})$. Let us
[d[define]d] introduce for our purposes the notation ${\cal O}(2)$
to describe terms that are quadratic in $\chi^\mu,\,\overline{\cal
H}_\nu$. Noticing that for $G:= {\chi} \overline{\cal H}$ and
arbitrary functionals $\alpha^\mu$ and $\beta_\nu$, one has
$$
\{\alpha\overline{\cal H} + \beta\chi,\, G\} = 
(\alpha + \beta\{\chi,\,\chi\})\overline{\cal H} - \beta\chi
+ {\cal O}(2)\,,
$$
(where $\beta\{\chi,\,\chi\}\overline{\cal H}$ must be interpreted
here and in similar expressions in the following as
$\beta_\nu\{\chi^\nu,\,\chi^\mu\}\overline{\cal H}_\mu$) one easily
obtains
$$
\{\Phi,\, G\}_{(2n+1)} = \{\Phi,\,\chi\}\overline{\cal H} 
+ \{\Phi,\,\overline{\cal H}\}\chi
+ {\cal O}(2)\,,
$$
for $n=0,1,2,3,\ldots$\ , and
$$
\{\Phi,\, G\}_{(2n)} = \Big(\{\Phi,\,\chi\}
+ \{\Phi,\,\overline{\cal H}\}\{\chi,\,\chi\}\Big)\overline{\cal H}
- \{\Phi,\,\overline{\cal H}\}\chi + {\cal O}(2)\,,
$$
for $n=1,2,3,\ldots\ .$ With these results, using the expansion
(\ref{defK}) we find
\bea {\cal K}_{\Phi} &=& \Phi +
(1-e^{-\lambda}) \{\Phi,\,\overline{\cal H}\}\chi \nonumber
\\ &+& e^{\lambda}\Big( (1-e^{-\lambda}) \{\Phi,\,\chi\}  \nonumber
\\ &+&     \frac{(1-e^{-\lambda})^2}{2} \{\Phi,\,\overline{\cal H}\}
\{\chi,\,\chi\}\Big)\overline{\cal H}+ {\cal O}(2)\,. \label{blow}
\eea Note in expression (\ref{blow}) that the off shell terms
diverge for $\lambda\to\infty$; this means in particular that the
limit ${\cal K}_{\Phi}\to {\cal I}_{\Phi}$ for $\lambda\to\infty$
can only be taken on shell. Let us define, with (\ref{theK}) in
mind, \bea\tilde{\cal K}_{\Phi} &:=&
\sum_{n=0}^{\infty}\frac{1}{n!}(1-e^{-\lambda})^n \chi^n \{ \Phi ,
\,
 \overline{\cal H}\}_{(n)}\nonumber \\ &=& \Phi + (1-e^{-\lambda}) \{\Phi,\,\overline{\cal H}\}\chi
+ {\cal O}(2)\,,
\eea
so finally we have
\beq
\tilde{\cal K}_{\Phi}=
{\cal K}_{\Phi}-e^{\lambda}\gamma_{{}_{\!\Phi}}\overline{\cal H} + {\cal O}(2)\,,
\eeq
with $$\gamma_{{}_{\!\Phi}} :=  (1-e^{-\lambda})  \{\Phi,\,\chi\}
+ \frac{(1-e^{-\lambda})^2}{2} \{\Phi,\,\overline{\cal H}\} \{\chi,\,\chi\}\,.$$
The functionals $\gamma_{{}_{\!\Phi}}$ carry the information of the lowest order off shell terms for
${\cal K}_{\Phi}$.
Now we can compute
\bea
\{\tilde{\cal K}_{\Phi^A},\,\tilde{\cal K}_{\Phi^B}\} &=&\{{\cal K}_{\Phi^A},\,{\cal K}_{\Phi^B}\}
- e^{\lambda}\{{\cal K}_{\Phi^A},\,\overline{\cal H}\}\gamma_{{}_{\!\Phi^B}} \nonumber \\
&-& e^{\lambda}\gamma_{{}_{\!\Phi^A}}\{\overline{\cal H},\,{\cal K}_{\Phi^B}\}
+ {\cal O}(\overline{\cal H},\,\chi)
\nonumber \\
&=&{\cal K}_{\{\Phi^A,\,\Phi^B\}}
- e^{\lambda}\{{\cal K}_{\Phi^A},\,\overline{\cal H}\}\gamma_{{}_{\!\Phi^B}}\nonumber \\
&-& e^{\lambda}\gamma_{{}_{\!\Phi^A}}\{\overline{\cal H},\,{\cal K}_{\Phi^B}\}
+ {\cal O}(\overline{\cal H},\,\chi)\,,\nonumber
\eea
and when we go on shell ($\overline{\cal H}\approx 0$), using (\ref{kh}),
\bea
\{\tilde{\cal K}_{\Phi^A},\,\tilde{\cal K}_{\Phi^B}\}&\approx& {\cal K}_{\{\Phi^A,\,\Phi^B\}}
-  {\cal K}_{\{\Phi^A,\,\overline{\cal H}\}}\gamma_{{}_{\!\Phi^B}} \nonumber \\
&-& \gamma_{{}_{\!\Phi^A}} {\cal K}_{\{\overline{\cal H},\,\Phi^B\}}+ {\cal O}(\chi)\,.\nonumber
\eea

Next we can take the limit $\lambda\rightarrow\infty$ and we obtain
\bea \{{\cal I}_{\Phi^A},\,{\cal I}_{\Phi^B}\}&\approx& {\cal
I}_{\{\Phi^A,\,\Phi^B\}} -  {\cal I}_{\{\Phi^A,\,\overline{\cal
H}\}}(\lim_{\lambda\to\infty}{\gamma}_{{}_{\!\Phi^B}}) \nonumber \\
&-& (\lim_{\lambda\to\infty}\gamma_{{}_{\!\Phi^A}}) {\cal
I}_{\{\overline{\cal H},\,\Phi^B\}} + {\cal O}(\chi)\,,\nonumber
\eea which explicitly shows the role played by the off shell terms
of ${\cal K}_{\Phi}$ in the computation of $\{{\cal
I}_{\Phi^A},\,{\cal I}_{\Phi^B}\}$. These terms will gently conspire
to bring the Dirac bracket on stage. Indeed, taking the limit
$p\rightarrow p_{{}_{\!G}}$, which is $\chi\rightarrow 0$, we obtain
\bea \{{\cal I}_{\Phi^A},\,{\cal
I}_{\Phi^B}\}{}_{\big|_{p_{{}_{\!G}}}}&=&
\{ \Phi^A,\,\Phi^B\} \nonumber \\
&-& \{\Phi^A,\,\overline{\cal H}\}\Big(   \{\Phi^B,\,\chi\}
+ \frac{1}{2} \{\chi,\,\chi\}\{\overline{\cal H},\,\Phi^B\} \Big) \nonumber \\
&-& \Big(   \{\Phi^A\,\chi\}
+ \frac{1}{2} \{\Phi^A\,\overline{\cal H}\} \{\chi,\,\chi\}\Big) \{\overline{\cal H},\,\Phi^B\}
\nonumber \\&=&\{ \Phi^A,\,\Phi^B\}^*{}_{\big|_{p_{{}_{\!G}}}}
={\cal I}_{\{\Phi^A,\,\Phi^B\}^*}{}_{\big|_{p_{{}_{\!G}}}}
\,,
\eea
which is (\ref{DB2}).

\subsubsection{Extension to all fields  including lapse and shift }
\label{allfields} Up to this point we have
restricted the generic field $\Phi$ on which we operate to be other
than the lapse and shift. When $\Phi$ is any generic field, or functional of
the fields, the generator of gauge transformations must be taken in
its full form (\ref{thegen}), or in its equivalent form
(\ref{gengen}) obtained by the use of a special basis for the
constraints, ${\overline \zeta}_{(i) \alpha}=(\overline{ {\overline {\cal H}}}_\nu,\,{\overline
P}_\mu)$. This is the form
that interests us.

By its construction, see subsection \ref{anyfield}, this new basis
has the property that $\{\chi^{(i) \alpha},\,{\overline \zeta}_{(j)
\beta}\} \approx - \delta^i_j \delta^\alpha_\beta$, where $\chi^{(i)
\alpha}= (\chi^\mu,\, \dot\chi^\nu )$ are the
secondary and primary gauge fixing constraints. A bonus of this
construction is that the constraints ${\overline \zeta}_{(i)
\alpha}$ have strongly vanishing Poisson brackets among themselves.
Thus all the properties that have allowed us to obtain results like
(\ref{DB2}) or (\ref{theK}) hold with the only change being the
doubling of the set of constraints involved: Instead of
${\overline {\cal H}}_\nu$ now we must take ${\overline P}_\mu,\,\
\overline{{\overline {\cal H}}}_\nu$, and instead of $\chi^\mu$ now
we must take $\dot\chi^\mu,\, \chi^\mu$. With this simple
consideration, all results are extended to any field. In particular
the connection between Poisson brackets for the invariants and Dirac
brackets - now defined with a set of $16$ second class constraints -
for the associated fields, and also the obtention of the invariants
${\cal I}_{\Phi}$ as limits of canonical maps ${\cal K}_{\Phi}$, for
any field $\Phi$, without restrictions.

\section{Conclusions}

In 1955 P. G. Bergmann, in a plenary talk in Bern celebrating
``Fifty Years of Relativity Theory'', expressed the belief that
``genuine invariants would reveal themselves as extremely
complicated functionals of the presently known field quantities''
\cite{bergmann56}.  Nearly forty years later Torre confirmed
Bergmann's belief with a proof that in generic general relativity no
observables exist that can be written as spatial integrals of Cauchy
data and finite derivatives thereof \cite{Torre:1993fq}. In this
paper we have explicitly displayed generic invariants as series
involving derivatives of Cauchy data in principle  up to infinite
order. Others, in particular Dittrich and Thiemann, have defined
invariants as formal power series. However, we are able to establish
a relation between functional invariants and specific gauge choices.
Equivalently, we have shown that invariants are obtained through a
choice of intrinsic coordinates.  The construction of invariants and
the demonstration of the equivalence of the two points of view is
achieved through the use of the underlying canonical diffeomorphism
symmetry group of generally covariant theories. We should point out that the observables that are obtained through these constructions are of course not functionally independent.
Given  the two degrees of freedom of pure gravity, there are in phase space four,
or rather $4(\times 3 \times \infty)$ functionally independent
observables. The proofs of 
our results are local in the sense that they can be 
applied to a region of spacetime. Global issues are not adressed 
in the present formulation.

We expect that our contribution will help to clarify some controversial issues that are still debated in the literature regarding the notion of gravitational observable. We cite in particular the  "frozen time" issue. We identify as a fundamentalt origin of many of these misunderstandings the fact that different authors do not use the same definition for common words like "gauge transformation". In Appendix A we have tried, in the guise of an informal dialogue, to pinpoint the most common causes of misunderstanding.  We do not, of course,  by any means claim to have pronounced the final words on any of them. The fact that our proofs are local means also that we are still far from a final and comprehensive description of the whole picture.

We have established in this paper a broad geometrical interpretation
of the construction of observables in generally covariant theories.
In particular, we have argued that there exists two basic equivalent
points of view as regards the construction of observables once the
solutions of the equations of motion are given.
\begin{itemize}
\item The first  point of view, and the one that enjoyed
particular emphasis, relies on the existence of a genuine
diffeomorphism-induced canonical gauge symmetry group. This group
realizes as active canonical transformations all changes of
canonical variables that result from general changes of spacetime
coordinates. We identify the group as ``diffeomorphism-induced''
because the resulting transformations depend on the functional form
of some or all of the components of the metric field. Indeed, in
order to be able to implement the transformation group, the lapse
and shift must be retained as canonical phase space variables, and
permissible diffeomorphisms depend on them in a compulsory manner.
We have shown that this group may be employed to construct functions
that are invariant under its action. The strategy is to choose an
explicitly spacetime coordinate dependent gauge condition, and then
to find the finite gauge transformation that transforms the fields
to that location on the gauge orbit where the gauge condition is
satisfied. The application of this finite transformation to all
field variables produces invariants associated with each and every
one of them. In order for this program to succeed it is mandatory
that the fields\footnote{These fields may be independent
fields or functionals of other fields. An example of the second case
is the use by Komar and Bergmann \cite{berg-kom60} of Weyl scalars,
also considered in \cite{ps04}.} $X^\mu$, those that are set equal to the
coordinates $x^\mu$ in the gauge fixing procedure, transform under
general coordinate transformations as spacetime scalars.

\item This brings us to the second equivalent view. The gauge choice
is nothing other than the selection of that system of spacetime
coordinates for which the fields $X^\mu$ produce the results
$X^\mu(x) = x^\mu$. This means that we are choosing the values of these 
scalar fields as the coordinatization of the spacetime. Users sitting at 
different points $p$ on the gauge orbits have phase space solutions $\Phi_p$ with distinct
functional forms. Each is given explicit instructions on developing
a potentially infinite series in powers of their coordinates
$x^\mu$, namely 
\bea 
&&{\cal I}_\phi \approx \nonumber\\ 
&& \sum_{n_\mu=0}^{\infty}
\frac{1}{n_0!\,n_1!\,n_2!\,n_3!}\,
(x^0)^{n_0}(x^1)^{n_1}(x^2)^{n_2}(x^3)^{n_3}\, {\cal
C}_{n_0,n_1,n_2,n_3}\ , \nonumber 
\eea
 with coefficients 
\bea 
&&{\cal C}_{n_0,n_1,n_2,n_3}:= \nonumber\\
&&{\cal I}_{\{\{\{\{ \phi_p,\,,\overline{\cal
H}_{0\,p} \}_{(n_0)}\,,\overline{\cal
H}_{1\,p}\}_{(n_1)}\,,\overline{\cal
H}_{2\,p}\}_{(n_2)}\,,\overline{\cal
H}_{3\,p}\}_{(n_3)}}{}_{\!\Big\vert_{x^\mu=\,0}}\,. \nonumber
\eea
  (Where setting $x^\mu=\,0$ refers only to the explicit dependencies.) 
These coefficients are constant; they are invariant
under under diffeomorphism-induced canonical transformations, i.e.,
under displacement from $p$ to $p'$ along the gauge orbit.
Furthermore, we have shown that these constants are nothing other
than the derivatives in a Taylor expansion of solutions at the
gauge-fixed location $p_G$ on the gauge orbit.  Notice that if the user works with the new fields
${\hat\Phi}^A = {\cal I}_{\phi^A}$ as the fields of her spacetime,
her new description is exactly as if her original coordinates played
the role of the intrisinc coordinates.

\item

We have also proven that the Poisson brackets of the invariants
${\cal I}_\phi$ are identical with the invariant  associated with
Dirac brackets, i.e., $\{{\cal I}_{\Phi^A},\,{\cal I}_{\Phi^B}\}
\approx {\cal I}_{\{\Phi^A,\,\Phi^B\}^*}$. This equality holds for
all the canonical variables, including the lapse and shift.

We are able to express all the invariants ${\cal I}_\phi$ 
as limits of canonical maps applied to the original fields. This aspect 
throws new light on the emergence of the Dirac bracket just mentioned. On the other 
hand, out of our invariants, which satisfy the EOM, one can obtain constants of 
motion with no explicit time dependence. These constants of motion, which are obviously 
observables, albeit of another kind, are generators of rigid Noether symmetries.

\item
Our results on the Dirac bracket connect our findings with Dirac's procedure 
for establishing the strong vanishing of Poisson brackets of constraints
and gauge conditions through the introduction of Dirac brackets. Of
course one must choose gauge conditions of the form $x^\mu = X^\mu$,
where the $X^\mu$ are four suitably chosen spacetime scalar
functions of the canonical variables. To our knowledge we are the
first to establish this detailed connection for the full set of
canonical variables.  Dirac had originally introduced this method as
a way of eliminating inconsistencies in passing from the classical
theory to the quantum theory. But since he lacked a geometrical
interpretation of the resulting formalism, he and others who
followed his lead tended to focus almost exclusively on attempting
to identify a minimal complete set of invariants. In particular, the
lapse and shift were simply eliminated from the formalism. This was
and remains a mistake when passing to the quantum theory. The Dirac
bracket of the lapse and shift carries physical information. The
quantum non-commutativity of lapse and shift with the remaining
quantum observables is an outcome of the specific choice of
intrinsic space and time coordinates.  As a consequence of this
choice the full metric in the quantum theory will be subject to
fluctuation - yielding a quantum ``thickening'' of the light cone in
an appropriate semi-classical limit.

\item We think that our approach makes a deep connection with the 
``evolving constants of motion" program. In particular, the elucidation 
of the different roles of the explicit and implicit - i.e., through the 
fields - time dependences in the observables proves to be a key 
ingredient in the full conceptual clarification of this program.

\end{itemize}
We wish to stress one additional aspect of our construction of
invariants: They are obviously solutions of Einstein's equations, and they make use of a set of selected scalars which define the 
intrinsic coordinates. It
is in this respect that we detect a potential disagreement with the
program of partial and complete observables that has been advanced
by Rovelli \cite{rovelli02}, and further elaborated in the canonical
framework by Dittrich \cite{dittrich07} and Thiemann
\cite{thiemann06}. Only if the partial variable is a spacetime
scalar will their construction of complete observables correspond to
an acceptable gauge fixing. In other words, some choices of partial observable as
coordinate time might not be legitimate gauge choices. 

\vspace{4mm}

Finally, we would like to express our most respectful admiration for
the work of Peter Bergmann as regards the concept of observables in
general relativity. He has inspired us as a teacher and friend. 
Many of our findings were conceptually anticipated in his short
review paper \cite{bergmann61}. There we find the idea that, when
considered as symmetry generators, the constants of motion obtained
from the observables take the configurations out of  the gauge orbit. 
He also anticipated the distinction
between the explicit time dependence of the invariants, necessary to
enforce the satisfaction of the equations of motion, and the
implicit one, which we discuss in section \ref{getnoether}. Again
with the advantage of hindsight one can anticipate from his
considerations the ``evolving constants of motion" program. He even
outlined what should be the role of the Dirac bracket in the algebra
of invariants, although without the symmetry group theoretical
foundations that we have established in this paper.

\section*{Acknowledgments}
JMP acknowledges partial support form  MCYT FPA 2007-66665, CIRIT GC
2005SGR-00564, Spanish Consolider-Ingenio 2010 Programme CPAN
(CSD2007-00042). He also thanks the Max Planck Institute for the
History of Science and the Theoretical Physics Group at Imperial
College London for their warm hospitality at different stages of
this work. DCS thanks both Austin College and the Max Planck
Institute for the History of Science for their generous support.

\section*{Appendix A. A dialog on canonical gravity, gauge symmetry and dynamics}

 In this appendix we attempt to communicate, in the form of dialogue, 
our views on some subjects that are still controversial or have been a 
source of misunderstandings in the canonical formulation of gravity. 
The interchange is between two subjects, A. and B., the latter representing our point of view. We will essentially touch upon three issues: the gauge group of canonical gravity, the meaning(s) of gauge 
transformations, and finally the infamous {\sl ``time is frozen, nothing happens"}
problem in canonical gravity. 
We progress from very na\"{\i}ve misunderstandings, which have by now been largely clarified, to some confusions which still persist in the literature. 
What is said concerning canonical gravity can be extended in 
obvious ways to other generally covariant theories. 

\subsection*{The gauge group of canonical gravity}

 A.: Sometimes I ask myself why Einstein's theory, which has such an aesthetic
appearance in the Lagrange-formulation becomes so ugly-looking in
its Hamiltonian-form. Beyond that, being based upon a $3+1$ decomposition, it is quite clear that 
canonical gravity is not able to describe the full $4$-diffeomorphism invariance of the Lagrangian 
formulation of general relativity. Having committed to a given $3+1$ decomposition means that a partial 
gauge fixing is in effect since the diffeomorphisms that do not preserve the foliation must be 
excluded. 

B.: Let us skip matters aesthetic, and cut to the chase. I profoundly disagree with the last thing you 
said. Nothing prevents diffeomorphisms from acting because the gauge group must be understood as a group of active 
diffeomorphisms and, as such, it never change the foliation.

A.: But it is clear that a diffeomorphism generated by a vector field $v^\mu$ will change the foliation as 
long as $v^0$ depends on the spatial coordinates.

B.: This is true in the passive view of diffeomorphism invariance, but we are interested in a canonical 
realization of the gauge group, that is, with generators acting through the Poisson brackets. This is an 
{\sl active} action in the sense that it modifies the field configurations but leaves unchanged the 
coordinates. Active and passive views must be neatly distinguished.

A.: I am happy to concede, but then let me mention what I see as 
a problem with the canonical realization of the gauge group.
Assume that it is possible to treat everything infinitesimally, i.e.
near the group-identity. We know for instance that in case of an
infinitesimal coordinate transformation $\delta x^\mu=
\hat{x}^\mu-x^\mu= {\rr - }\epsilon^\mu(x)$ the infinitesimal variation of a
tensorial object T is given by the Lie derivative $\pounds_{\!\epsilon}
T$. (We should leave out further complications due to the presence
of spinorial fields in this discussion.) But here things already get
hard, since the diffeomorphism group is more complicated than a
finite dimensional Lie group. This is for instance reflected in the
Poisson-bracket structure of the Hamiltonian and momentum
constraints in canonical gravity. You don't have structure constants but 
structure functions. The diffeomorphism group is not realized in phase space.

B.: Not so fast! One must be very careful when moving into phase space. 
Let me first address some aspects of diffeomorphisms in configuration-velocity space. 
Bergmann and Komar observed in the early seventies that Einstein's
field equations (respectively, the Hilbert-Einstein action) are not only
invariant under point/contact transformations
$(\hat{x}^\mu=f^\mu(x^\nu))$, but also under transformations which
additionally may depend on the metric fields and their derivatives;
i.e. $(\hat{x}^\mu=f^\mu(x^\nu; g_{\mu\nu}(x),...)$. This is more
than a spacetime diffeomorphism in the usual sense of (passive)
coordinate transformations. Whereas the diffeomorphism group is
acting on the Riemannian manifold (locally describable in the passive 
view as general coordinate transformations), the larger group, which 
is certainly a diffeomorphims-induced gauge group, acts on the space of
metrics of the Riemannian manifold - and on every other field that is around.

A.: Why  make things even more complicated by investigating this
far larger group?

B.: The gauge group is what it is, not what you would like it to be.
There are many answers to your question, that is,  many ``becauses": Because an
important subgroup of this larger group can be realized in phase-space,
because this subgroup reveals the explicit form of the gauge
generators, because this subgroup gives a clue as to how to interpret even
a gravitational Hamiltonian as being responsible for unfolding
dynamics, and because this subgroup leads you to a better
understanding of observables.

A.: Wow, seems that the ``subgroup of the generalized symmetry group"
cures my headache - and not only mine. By the way, is this the
Bergmann-Komar group?

B.: Well, it depends on who you ask. Some authors mistakenly denote the full metric dependent group as the Bergmann-Komar group. We reserve this name for the projectable subgroup. OK, now you can follow me as we explore the fate of symmetries in going from the configuration-velocity space to phase space.

A.: I think we can skip this, since this Dirac-procedure is already
standard and described in textbooks. At the very end we arrive at a
extended Hamiltonian, being the sum of the canonical Hamiltonian
and arbitrary linear combinations of the first-class constraints.

B.: Again you are going too fast. Since I know some German: ``Soviel Zeit
mu{\ss} sein" - to not forget the contribution of Peter Bergmann and
his collaborators (and also that of L. Rosenfeld - but this is
another story). But what is more essential for our discussion: We
know that because of the singular character of the GR-Lagrangian the
Legendre transformation from the configuration-velocity space to the
phase space is not invertible.

A.: Sorry to interrupt you again, but to arrive at a Hamiltonian even in this
singular situation  is exactly the task of the
Dirac-Bergmann-algorithm.

B.: The algorithm is one thing, but understanding the input and the
output of the procedure is another story.

A.: The input is the Lagrangian ...

B.: ... with its symmetries. Just wondering - according to you which generalized diffeomorphism symmetry survives the Legendre transformation?

A.: Is it the Bergmann-Komar-group?

B.: You are very clever! Neither the diffeomorphism group nor the
larger general field-dependent group allow for a transition from the
tangent-space to the cotangent-space. 
In order to be
Legendre-projectable the field-dependent group must be restricted in
a specific way, already specified in the 1972 article by Bergmann
and Komar.

A.: Could you make this more precise?

B.:  Legendre-projectability restricts the functional form of the
$\epsilon^\mu(x^\nu, g_{\varrho\sigma}, ...)$ to
$$\epsilon^\mu(x^\nu, g_{\varrho\sigma}, ...)= n^\mu \xi^0 + \delta^\mu_a
\xi^a,$$ where the $\xi^\mu$ are descriptors depending only on the
three-geometry (three-metric components and their spatial
derivatives) and $n^\mu$ is the normal to the $t=\textrm{const}$
hypersurface, expressed by the lapse $N^0=N$ and the shift functions
$N^a$ as
$$n^\mu = \{N^{-1}, -N^{-1} N^a \}.$$

A.: To make things easier let us take the by now standard gauge
choice $N=1$ and $N^a=0$. We know, and this is already text-book
knowledge that ``lapse and shift should not be viewed as dynamical
variables".

B.: Careful! Through an untimely ``so-called" gauge choice you are
loosing insights in the structure of the phase-space version of
gravity: If you fix the lapse and the shift, you are no longer able
to identify the ``so-called" gauge generators, nor are you able to
recognize the difference between the ``so-called" gauge generators
and the Hamiltonian. And yet you are also worried about the fate of
general covariance in going from the Lagrangian to the Hamiltonian
in gravitational theories. It turns out, and this is important, that in order to see the diffeomorphism-induced symmetry in phase space one is forced to treat the
lapse and the shift functions in the theory as genuine fields.

A.: Okay, I accept that - however with a grain of salt, since I learned
that the lapse and the shift functions are devoid of any physical
meaning. By the way, why are you so insistent  in talking about
``so-called" gauge-blah-blah?

B.: I'm doing this, because it seems that when considering canonical gravity
there seem to be at least two different understandings of what ``gauge" means.
 
A.: Why so? One has the diffeomorhisms...

B.: Which diffeomorphisms do you have in mind here? Automorphic
mappings of manifolds, generalized symmetries in the sense of
Bergmann and Komar, diffeomorphism-induced transformations in
phase-space, or perhaps, following Dirac, gauge transformations at a fixed time?

\subsection*{The meaning(s) of gauge transformations}

A.: Okay, you got me. Seems it is time, to get deeper into the
meaning of gauge symmetries. When Dirac wrote his book, 
what he had in mind briefly as
follows ... wait ... there is a nice description in Rovelli's book \cite{Rovelli:2004tv}:
"\textit{Consider a system of evolution equations in an evolution
parameter t. The system is said to be ``gauge" invariant if evolution
is under-determined, that is, if there are two solutions that are
equal for $t$ less than a certain $t_0$.}"

B.: This is a possible presentation of the ``gauge symmetry" phenomenology, but one must 
proceed with extreme care as regards the definitions. 
The original diffeomorphism-symmetry maps complete solutions of the field
equations to other solutions, but if you read carefully Dirac's book on constrained systems, 
you will see that when he discusses gauge transformations, he refers to a fixed time,
namely $t_0$. What makes us believe that Dirac's notion of gauge
invariance is the same as the gauge invariance considered by Bergmann,
which is the one that maps solutions of the EOM into solutions? I observe that in the community
there is no clear distinction among these two notions. And thus
there is no clear distinction about what a ``gauge generator" is
meant to be. This disagreement underlies the famous dispute over whether dynamics is ``frozen'' in generally covariant theories like general relativity.

A.: But we know from Dirac's work that all first-class constraints
generate gauge transformations.

B. Regretfully Dirac's approach to gauge transformations has caused 
a lot of misunderstandings. His concept of a gauge transformation was not that
of mapping solutions of the EOM into new solutions. He worked at a fixed 
time - the evolutionary parameter - and so his concept was rather that of 
relating two sets of initial conditions - at that given time - that respectively 
belong to two gauge equivalent solutions. 
 
A.: So, what is the difference in Dirac's understanding of ``gauge transformation"
and Bergmann's notion?

B.: That's easy to state: Dirac's ``gauge-transformations", generated by all
first-class constraints (well, if certain mathematical regularity conditions hold), 
are valid for a fixed time only - which can be taken as the time  at which initial conditions are formulated. But when one considers all possible times, which is 
necessary if we want to act on a whole solution of the EOM, then the gauge generators, 
as Bergmann pointed out, are specific combination of first-class constraints, with a certain number of
arbitrary functions and  their time derivatives  attached to these 
constraints. 

In fact you don't even need to consider general covariance to grasp the distinction 
between Dirac's and Bergmann's conceptions. Just take pure Maxwell theory with gauge 
field $A_\mu$. There is a primary first class constraint, namely the momentum conjugate to $A_0$, 
and a secondary first class constraint, the Gauss constraint. To generate the gauge 
transformation $\delta A_\mu = \partial_\mu \Lambda$ ($\Lambda$ is an arbitrary function) in phase space, 
you need to construct a gauge generator made with a specific combination of the two constraints, with coefficients $\Lambda$ for the secondary one and $\dot\Lambda$ for the primary one. This is Bergmann conception. 
Of course if you consider just a fixed time $t_0$, since $\Lambda$ is an arbitrary function, $\Lambda$ and $\dot\Lambda$ become independent functions of the spatial coordinates, and that is why in Dirac's view, both constraints generate gauge transformations, but we must insist that this last picture is only valid at a 
fixed time! You can see with your own eyes that neither of these constraints alone generates transformations mapping solutions into solutions.

In the case of GR, the gauge generators are explicitly
$$G_{{\xi }}(t) =    P_{\mu} \dot\xi^{\mu} + ( {\cal H}_{\mu}
+ N^{\rho} C^{\nu}_{\mu \rho} P_{\nu}) \xi^{\mu}.$$ Here the ${\cal
H}_{\mu}$ are the well-known Hamiltonian and momentum-constraints,
and the $C^{\nu}_{\mu \rho}$ the structure coefficients in their
PB-algebra (called Dirac-algebra by some). The $P_{\mu}$ are the
momenta canonically conjugate to the lapse and shift-functions
$N^{\mu}$, and a spatial integration over repeated indices is to be
understood. The $\xi^{\mu}$ are arbitrary functions of the spacetime coordinates 
as well of the field components except for the lapse and shift. 

A.: Again I see the lapse and the shift function in this expression.
Things would become easier for the gauge choice $N=1$, $N^a=0$.

B.: Yes, things would become easier for some explicit calculations,
however not for the interpretation of the gauge generators as
generating exactly what they are supposed to generate as symmetry
operators, namely for any object $\phi$ in the theory
$$\pounds_{\!\xi} \phi =\{\phi, G_{{\xi }}\}.$$
Note that this Poisson bracket is an equal time bracket. In order to 
construct the full gauge transformation, mapping solutions into solutions, 
one needs to consider all times (or at least a finite interval for the time 
parameter).

\subsection*{Gauge transformations versus dynamical evolution}

A.:  Even though it's been hard to follow you with so many different notions of 
gauge invariance, I think I've finally got you: 
The Dirac-Hamiltonian for a generally covariant theory is known to
be $$H_D=N^\mu{\cal H}_{\mu}+\lambda^\mu P_\mu.$$ ($\lambda^\mu$ are the 
arbitrary functions of the dynamics) Thus the choice
$\xi^\mu=N^\mu$ leads to a gauge generator $G_N$ once you take into
account the equations of motion $\dot{N}^\mu=\lambda^\mu $. Thus I make 
the strong claim that the Hamiltonian is a specific gauge generator. 
And if the Hamiltonian is a gauge generator (even in the
sense of Bergmann), how can you escape from interpreting this as
leading to ``frozen-time"? There is no dynamics at all!

B.:  Here is the quick and easy response. The generator $\delta t (N^\mu{\cal H}_{\mu}+\dot N^\mu P_\mu)$ does serve to replace solutions at time $t$ by the original solutions evaluated at $t-\delta t$. But it performs this function only on one particular member of each equivalence class of solutions, namely those for which the lapse and shift are the chosen explicit function $N^\mu$. On all other members of equivalence classes the effect is to generate variations that are distinct from global translations in time.   

But let me try to
convince you by looking  more closely at the geometry and the transformations we
are talking about. For this purpose I will denote the space of fields
obeying the GR field equations - one can include matter fields as 
well - by ${\cal S}$; thus points in ${\cal S}$ are specific spacetimes 
with the fields - solutions of the EOM - described 
in a particular coordinatization. Consider the field content of a point
$p$ in ${\cal S}$. Let us focus on the data for the fields at time
$t_0$ and let us call $D$ these data. With a specific selection of
the arbitrary functions $\lambda$, there exists a Dirac Hamiltonian,
$H(t) = N^\mu{\cal H}_{\mu} + \lambda^\mu P_\mu$, which dictates,
through the Poisson brackets, the time evolution in $p$.
Particularly, for an infinitesimal $\delta t$, this Hamiltonian
tells what are to be the field data at the hypersurface labeled by
$t_0+\delta t$. Let us call these new field data $D'$. Now, if we do
this for all times $t$, the result is that of course we have
remained exactly at the same point $p$ in ${\cal S}$, because the
dynamics as described by a given observer, takes place within a given 
spacetime in a given coordinatization.

A.: Is this long  exposition meant to persuade me that the Hamiltonian determines the dynamical evolution in phase space?

B.:Well ... yes. But in addition I would like to point out to you the
the difference between a Hamiltonian and a gauge generator.  So let me go
on. Consider the gauge generator that, after an appropriate choice
of the descriptors, happens to coincide in its mathematical
expression with the Dirac Hamiltonian at time $t_0$. Due to this
coincidence, its action will of course transform the field data $D$
into $D'$, but these data $D'$ are now conceived at time $t_0$,
because the gauge transformations are equal-time actions. What
happens is that we have moved from $p$ to another gauge equivalent
spacetime $p'$. If we undertake the same procedure for any time $t$ (continuing to assume that the descriptors at time $t$ match up with the lapse and shift at time $t$)
we will end up having mapped the whole spacetime $p$ to $p'$. Notice
that the field configurations in $p$ and $p'$ just differ in the
time label, and that a passive diffeomorphism $t\rightarrow t-\delta
t$ will make both descriptions identical. Obviously this fact should
not be a surprise, but should be viewed as a simple consequence of
our fundamental understanding of spacetime gauge symmetry. Thus the
fact that the gauge generator can mimic the Hamiltonian has nothing
to do with the fact that there is real physical\footnote{See below
for clarifications on the meaning of "physics".} evolution in a
given spacetime $p$, where we may consider events, coincidences,
causal structure, observables, and so on. Dynamical evolution in $p$
is not gauge action on $p$.

A.: Sorry, you almost manage to confuse me, so let me use my own
language. On one side we have $D$ and $D'$ as field configurations
connected by a gauge transformation. On the other we know that, in
some spacetime, the configuration $D'$ lies in the future of $D$. 
Since by definition a gauge transformation does not change the physics, 
we deduce that the physics in $D$ and $D'$ are the same. So the future 
is gauge equivalent to the past and therefore "nothing happens". 
How do you address that?

B.: Let me remind you of what we discussed earlier: that in 
generally covariant theories we must
distinguish two notions of gauge transformation, namely the ones we
previously called by the names of Dirac and Bergmann, respectively.
Since the symmetry-inspired (Bergmann) notion is about mapping
solutions of the equations of motion to solutions, we need to have
entire field configurations, not just configurations at a given time
$t_0$, as  occurs with $D$ and $D'$. In saying that $D$ and $D'$
are connected by a gauge transformation you inadvertently changed
the concept of gauge transformation - from Bergmann's to Dirac's - but 
intended to keep intact its intepretation. That is a mistake.

A.:  Just to get your point: Are you saying that the phrase ``A gauge
transformation does not change the physics" is wrong?

B.: It depends on what you mean by ``the physics" and by a ``gauge
transformation". Note that $D$ and
$D'$ can be conceived as settings of initial conditions. The fact
that these two sets of initial conditions are related by a gauge
transformation {\sl effected at a given time $t_0$} means that both
$D$ and $D'$ are good data to build, using the dynamics, the same
physics. This is what we have seen before with the spacetimes $p$
and $p'$, which are gauge equivalent. With the word ``physics" here
we mean the entire spacetime, with the entire history, modulo gauge
transformations. We may call this ``physics" the {\sl ``entire-physics"}.
This is the ``physics" that enters in your sentence ``a gauge
transformation does not change the physics", because it refers to
mapping solutions into solutions. In this physics the statement is
true.

A.: This being said, I assume that there is another meaning of
``physics" in which the sentence ``A gauge transformation does not
change the physics" is indeed wrong.

B.: Yes indeed.  But you must take now Dirac's gauge transformations 
at a single time. As we have just seen, $D$ and $D'$ are related in 
this sense. It is obvious that if $D'$ is a fixed-time field
configuration in the future of $D$ (we may consider here finite time
separations instead of infinitesimal ones) in a given spacetime $p$,
both are equally good data from which one can reconstruct the entire
spacetime and so both belong to the same "entire-physics". But if we
prefer to stay in a more down-to-earth perspective, regarding
configurations at a given time (let us call it {\sl ``timeslice-physics"}) 
in a given spacetime, then of course the timeslice-physics in 
$D$ and in $D'$ can be very different, although the entire-physics 
is the same. Perhaps in $D$
you were not born yet and in $D'$ you were. That's a big difference,
and observable, isn't it?

\section*{Appendix B. From weakly to strongly vanishing Poisson brackets of first class
constraints.} 

In their monograph  \cite{henneauxt94} Henneaux and Teitelboim considered in
Chap.5.2 the idea of ``Abelianization of Constraints", an idea already 
present in another language in classical monographs on differential equations, like 
\cite{Eisenhart}. In this respect, a particularly efficient technique is that of Dittrich 
\cite{dittrich07} and Thiemann \cite{thiemann06}, which we
adopt in \ref{descrip}. We will prove that these ``Abelianized"
constraints lead to strongly vanishing first class constraints in
the sense of Dirac. In Dirac's terminology, a function $f$
strongly vanishes (denoted as $f\equiv 0$) in phase space if it
vanishes and in addition its differential also vanishes on the
constraint surface.

For simplicity, we shall use the language of mechanics. Consider a
$d$-dimensional manifold with $n$ ($n\leq \frac{d}{2}$) first class,
independent and effective ( A constraint is said to be effective it it has
a non vanishing differential on the constraint's surface.) constraints $\phi_i$. 
They define the
surface $\cal M$ and so they satisfy $\{\phi_i,\,\phi_j\}= f_{ij}^k
\phi_k$. Associated vector fields are
$$
V_i = \{-,\,\phi_i\}\,,
$$
so that
$$[V_i,\,V_j]\approx f_{ij}^k\,V_k\,,
$$
where the symbol $\approx$ means that the equality is valid on $\cal M$.

Now consider $n$ independent functions $F^i$ such that
$\det{V_i(F^j)}\neq 0$. Next take the functions $F^i$,$\phi_j$ and
$d-2n$ extra functions to make a change of coordinates in the phase-space, 
at least in a neighborhood of
$\cal M$. If the original coordinates -positions and momenta- where
$x$, we call the new coordinates $y$ so that $y^i = F^i(x),\ y^a =
F^a(x)$, where $a= n+1 \dots d$ and $F^a$ include the constraints
$\phi_i$.

Express the vector fields in the new coordinates
$$
V_i= V_i(F^j)\partial_{y^j} + V_i(F^a)\partial_{y^a}\,.
$$
Next define $B_i^j = (V_i(F^j))^{-1}$ amd make independent linear combinations of the vector fields
by defining $\bar V_i = B_i^j\,V_j$. It turns out that
$$\bar V_i = \partial_{y^i} + D_i^a \partial_{y^a}\,,$$
for some coefficients $D_i^a$. Since we just made a linear combination of the vector fields, they
still satisfy a closure property on ${\cal M}$,
$$[\bar V_i,\,\bar V_j]\approx \bar f_{ij}^k\,\bar V_k\,,
$$
but on the other hand, given the form of $\bar V_i$ in the new coordinates, it is clear that
$[\bar V_i,\,\bar V_j]$ can not have $\partial_{y^k}$ terms on the right hand side and therefore we can not write
$\bar V_k$
on the right hand side. We conclude that $[\bar V_i,\,\bar V_j]\approx 0$ (or equivalently, $\bar f_{ij}^k\approx 0$).

Now consider a change of basis for the constraints, along the same lines, that is, $\bar \phi_i
= B_i^j\,\phi_j$., then
$$\{-,\, \bar \phi_i\} = \bar V_i + {\cal O}(\phi) \,,$$
where by ${\cal O}(\phi)$ we mean vector fields that vanish on ${\cal M}$. We thus have
$$\{f,\,\{ \bar \phi_i,\, \bar \phi_j\}\}\approx [\bar V_i + {\cal O}(\phi),\,\bar V_j
+ {\cal O}(\phi)]f\approx [\bar V_i ,\,\bar V_j ]f\approx 0\,,$$
for any function $f$, which means that
$$\{ \bar \phi_i,\, \bar \phi_j\}= {\cal O}(\phi^2)\equiv 0\,.$$

\end{document}